\documentclass[11pt]{article}
\usepackage{graphicx}
\usepackage{amssymb}
\usepackage{amscd}
\usepackage{mathrsfs}
\usepackage{longtable,lscape}
\usepackage{amsthm}
\usepackage{amsfonts}
\usepackage{amsmath}
\usepackage{bbm}
\usepackage{float}
\usepackage{url}
\usepackage{csquotes}
\usepackage{wrapfig}

\usepackage{caption}
\usepackage{lineno}
\begin{document}
\title{\bf Heisenberg uncertainty principle and the strange physics of spaghetti}
\author{Massimiliano Sassoli de Bianchi\vspace{0.5 cm} \\ 
\normalsize\itshape
Center Leo Apostel for Interdisciplinary Studies, Brussels Free University, 1050 Brussels, Belgium \\ 
\normalsize
E-Mail: \url{msassoli@vub.ac.be}
\\ 
\normalsize\itshape
Laboratorio di Autoricerca di Base, 6917 Lugano, Switzerland \\
\normalsize
E-Mail: \url{autoricerca@gmail.com} \\
}
\date{}
\maketitle
\begin{abstract} 
\noindent This is an article written in a popular science style, in which I will explain: (1) the famous Heisenberg uncertainty principle, expressing the experimental incompatibility of certain properties of micro-physical entities; (2) the Compton effect, describing the interaction of an electromagnetic wave with a particle; (3) the reasons of Bohr's complementarity principle, which will be understood as a principle of incompatibility; (4) the Einstein, Podolski and Rosen reality (or existence) criterion, and its subsequent revisitation by Piron and Aerts; (4) the mysterious non-spatiality of the quantum entities of a microscopic nature, usually referred to as non-locality. This didactical text requires no particular technical knowledge to be read and understood, although the reader will have to do her/his part, as conceptually speaking the discussion can become at times a little subtle. The text has been written having in mind one of the objectives of the Center Leo Apostel for Interdisciplinary Studies (CLEA): that of a broad dissemination of scientific knowledge. However, as it also presents notions that are not generally well-known, or well-understood, among professional physicists, its reading may also be beneficial to them. 
\end{abstract}

\maketitle

\section{Introduction}

\noindent This article contains the (revised and slightly expanded) ``transliteration'' of a video that I published on YouTube on April 5, 2012, first in Italian \cite{SassoliSassoli2012YouTube-Italian} (my native language), then on August 27, 2012 also in English \cite{SassoliSassoli2012YouTube-English}. The Italian version received to date (June 15, 2018) more than 148,000 views, and considering that it is a video of almost an hour and a half, which is exclusively about physics, I consider it as an encouraging result. This also explains why at the time I decided to make the effort of producing an additional English version of the video, which however, probably due to my ``macaronic English,'' only obtained to date a little more than 18,000 views: a one order of magnitude difference with respect to the original Italian version. 

Regardless of the differences in terms of numbers of views, both the Italian and English videos received very positive comments (an event in itself quite rare on YouTube), which is the reason that led me to also write the present article. I hope in this way to do something pleasing to those who enjoyed the video, offering them the opportunity to retrace its contents in a form not only stylistically a bit more accurate, but also, perhaps, more suitable for the continuation of the reflection about its content. I also hope that this will allow the non-habitual users of YouTube to also access the explanations contained in the video, and that the present article version of it will be met with the same enthusiasm.

Before starting, let me bring back some of the positive comments I received in connection with the video. This not to indulge in some kind of narcissistic pleasure, but because these extemporaneous comments (here taken from the Italian version of the video) are able to express, I believe, some of the characteristics of the text that I hope you are in the process to read.

\begin{displayquote}
``Interesting and really well done. I deal with the philosophy of science in the USA; just a few hours ago I debated on the linguistic difficulty related to the sayability of the concepts of quantum mechanics in an unambiguous way and different from the classic ontology, which generates misunderstandings and often inaccuracies; I will report your valid presentation as an excellent example. Congratulations.''
\end{displayquote}

\begin{displayquote}
``Thank you for the explanation [...] I also liked the part about the spaghetti, a brilliant metaphor to understand the influence of the experimenter on the physical system, far beyond the banality of classical concepts.''
\end{displayquote}

\begin{displayquote}
``I find the explanation incredibly clear for those who want to have a general idea of the problems. The uncertainty principle drove me crazy because of its incompatibility with everyday experiences and it cut me off from a somewhat deeper reading on quantum physics. Thank you for giving us your time and your expertise!''
\end{displayquote}

\begin{displayquote}
``Really interesting. Thank you. But now I hate wooden cubes to death...''
\end{displayquote}

Of course, there were also some less enthusiastic comments, more critical about the content of the presentation and the way things were explained. Here is an example:

\begin{displayquote}
``Despite having a great (amateur) passion for the topic, after six minutes I got bored and lost... if you dive into charts and formulas, then the video is not for everyone. [...] If the things explained in the video I had read them in a book, it would have been the same.''
\end{displayquote}

Well, I hope that this last comment, although not very laudatory, portends a possible success also for the ``article format'' of my video-work. 

To conclude this brief introduction, I would like to say that most of what I will tell you in the following pages, is the result of an understanding that has developed thanks to the work of the so-called \emph{Geneva-Brussels school of quantum mechanics}, especially thanks to the research of \emph{Constantin Piron} (of whom I was the assistant in Geneva, for his famous course in quantum mechanics) and \emph{Diederik Aerts} (a student of Piron, with whom I have the pleasure today to collaborate).

Very well, I hope your reading will be enjoyable and thought provoking.

\section{A simple experiment}
\label{simple}

\begin{figure}[htbp] 
\begin{center} 
\includegraphics[width=9cm]{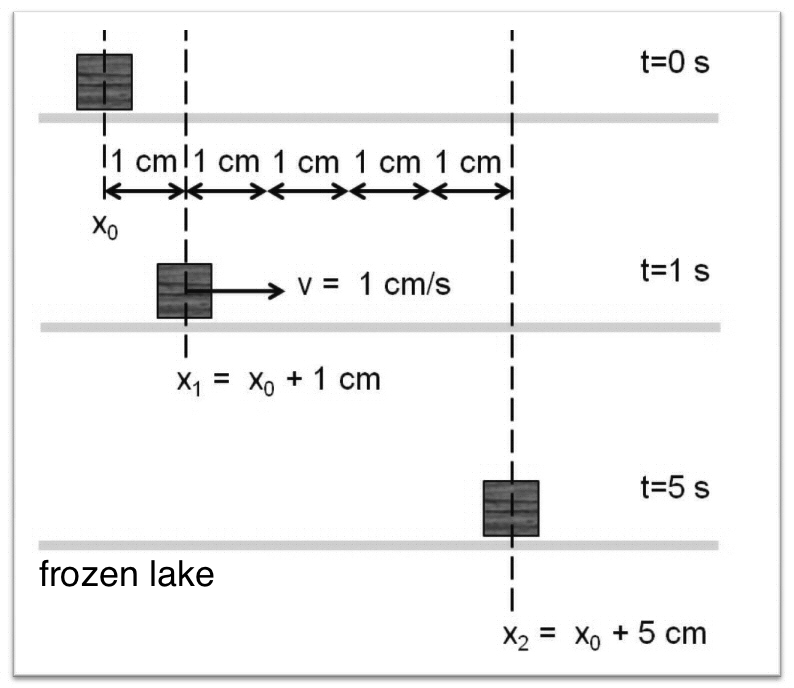} 
\caption{The three snapshots indicating the different positions of the wooden cube, at the three different instants of time $t=0,1,5$ seconds.} 
\label{Figure1} 
\end{center} 
\vspace{-0.5cm}
\end{figure}

\noindent Let me begin with a very simple experiment. We are on the surface of a frozen lake, by night. The physical system that we want to study is a wooden cube, and the instrument at our disposal to do so is a camera with flash. The experimental procedure is as follows: a colleague throws the cube on the ice, so that it will slide on the surface of the lake (without letting it rotate and assuming for simplicity that there is no friction). At this point, we take a first picture, at time $t = 0$~s. This first picture shows us that the wooden cube was, at that precise moment, in the position $x_0$ (see Figure~\ref{Figure1}). 

After exactly one second, that is, at time $t = 1$~s, we take a second picture. This second picture reveals to us that the cube was, at that precise moment, one centimeter away compared to the previous position, i.e., at the position $x_1 = x_0 + 1$~cm (see Figure~\ref{Figure1}). This allows us to conclude that the velocity $v$ of the cube is exactly one centimeter per second: $v=1$~cm/s.

To recapitulate, at time $t = 1$~s, we know both the position of the cube and its velocity. In other words, we jointly and simultaneously know the values of these two physical quantities. This enables us to predict with certainty any other position that the cube will occupy in later times. 

For example, given that we know that the cube moves with a velocity of 1 cm per second, we can predict with certainty that after further $4$ seconds, i.e., at time $t = 5$~s, it will be exactly in the position $x_2 = x_0 + 5$~cm, as is easy to confirm by taking one last picture, just at that moment (see Figure~\ref{Figure1}).

\section{Heisenberg uncertainty principle}

\noindent Let me consider now the famous \emph{uncertainty principle} (which should more properly be called, as it is the case for example in Italian, \emph{indetermination principle}) introduced in 1927 by the German physicist \emph{Werner Heisenberg} \cite{Heisenberg1927} (see Figure~\ref{Figure2}). 

\begin{figure}[htbp] 
\begin{center} 
\includegraphics[width=3cm]{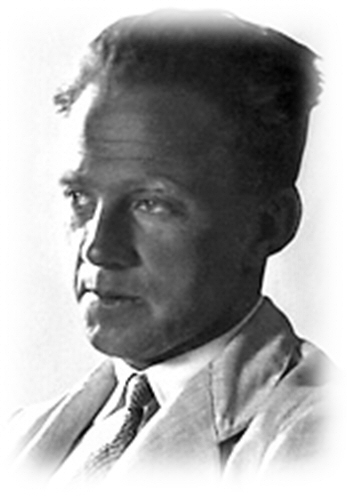} 
\caption{The German physicist Werner Heisenberg.} 
\label{Figure3} 
\vspace{-0.5cm}
\end{center} 
\end{figure}


What does this principle exactly tell us? Well, simply that, contrary to what we have just learned in relation to the wooden cube: 
\begin{displayquote}
\emph{There is no way to determine simultaneously, with arbitrary precision, both the position ($x$) and the velocity ($v$) of a microscopic particle, not even by using the most sophisticated measuring instrument.}
\end{displayquote}

There is of course no contradiction between this principle and our previous experiment, given that a wooden cube is not a microscopic entity, but rather a macroscopic one, i.e., a body of large dimensions. 

We can state \emph{Heisenberg uncertainty principle} (HUP) in a bit more precise way with the aid of a very simple mathematical relation. This relation states that the smallest error $Er(x)$ with which we can determine the position $x$ of a microscopic particle, at a given instant, multiplied by the smallest error $Er(v)$ with which we can determine, at the same instant, its velocity $v$, must always be, approximately, equal to a specific constant $c$. 

In mathematical language, what I have just stated is summarized in the following relation (the symbol ``$\approx$'' means ``approximately equal to''):
\begin{equation}
Er(x) \cdot Er(y)\approx c.
\label{HUP}
\end{equation}

To give an example, in the case of an \emph{electron}, if we measure the error on the position in centimeters (cm) and the error on the velocity in centimeters per second (cm/s), the value of the constant $c$ is approximately 1 centimeters squared per second (cm$^2$/s). 

To better understand the content of this relation, we can visualize it graphically, by representing it as a curve, so that only the points that lie on the curve satisfy the HUP (see Figure~\ref{Figure3}). Let us choose among them the point which is closest to the origin. As you can see, it corresponds to the situation where we have reduced the most, at the same time, both the position error and the velocity error. 

\begin{figure}[htbp] 
\begin{center} 
\includegraphics[width=6cm]{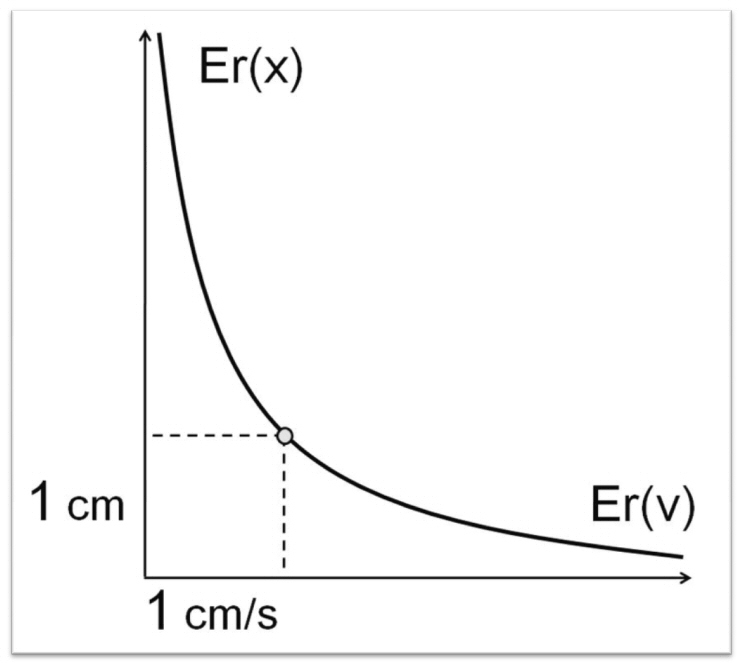} 
\caption{The points on the curve obey the HUP (here for an electron). The point which is evidenced, with coordinates $(1,1)$, is the one which reduces at best both the error on the position $x$ and on the velocity $v$.} 
\label{Figure3} 
\vspace{-0.5cm}
\end{center} 
\end{figure}

\begin{figure}[htbp] 
\begin{center} 
\includegraphics[width=6cm]{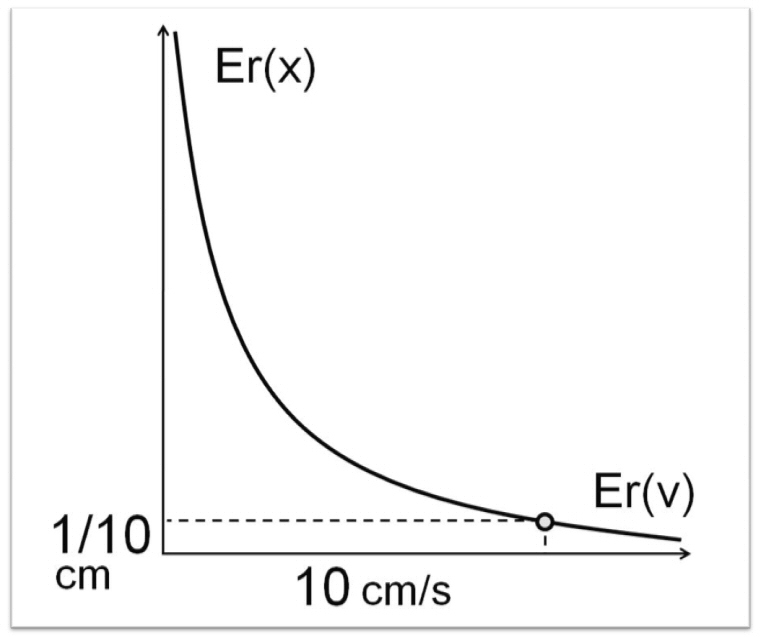} 
\caption{Reducing by one-tenth the error on the position causes the error on the velocity to increase by a factor of ten.} 
\label{Figure4} 
\end{center} 
\vspace{-0.5cm}
\end{figure}

But what if we wanted to further reduce, say of one-tenth, the error on the determination of the electron's position? To do so, and since we are forced to move on the curve, we must evidently slide the point to the right. But in doing so, while reducing the error on the position of one-tenth, at the same time we will increase by a factor of ten the error on the velocity (see Figure~\ref{Figure4}).

Same thing if instead of the position we try to reduce the error on the determination of the velocity: reducing by one-tenth the error on the velocity will cause the error on the position to increase by a factor of ten.

Very good, but let us now try to understand why Dr. Heisenberg invented, so to speak, his beautiful principle (which is not actually a true principle, as it can be deduced from more fundamental axioms of quantum theory). 

\begin{figure}[htbp] 
\begin{center} 
\includegraphics[width=6cm]{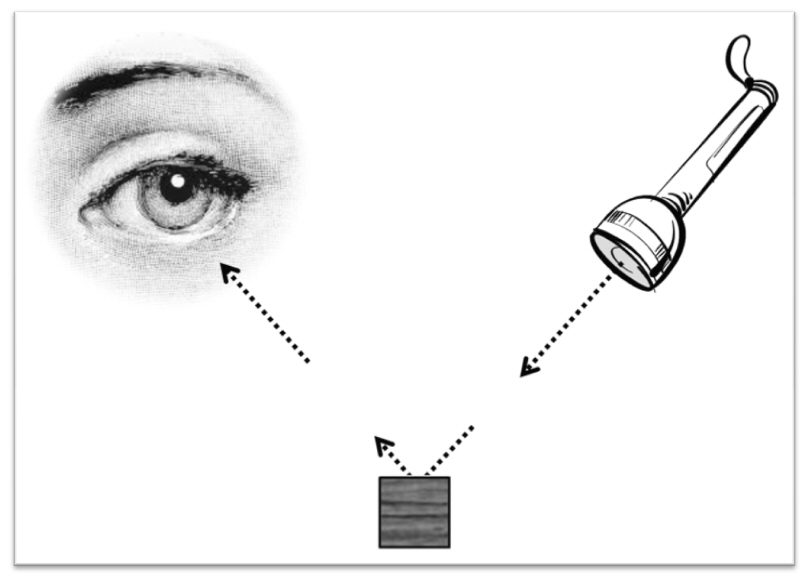} 
\caption{Seeing means observing scattered light.} 
\label{Figure5} 
\end{center} 
\vspace{-0.5cm}
\end{figure}

Let us first clarify what it means \emph{to see} a macroscopic body. Consider once again the wooden cube. If we want to see it, we must necessarily light it up with a light source, such as a flashlight. When the light rays strike the cube, they are deflected toward the detecting instrument, which in the present case is your eye, or better your eye-brain system (see Figure~\ref{Figure5}). 

In other words, to see an object means, roughly speaking, to detect the light coming from that object. We can observe that when the flashlight illuminates the cube, the latter is not in any way disturbed by it. This is therefore a \emph{non-invasive} observational process, through which we are able to \emph{discover} what already existed, regardless of our observation. 

Here we can discover not only the existence of the cube itself, but also its characteristics, like its shape and color, and of course its specific location in space. And as with the simple experiment of Sec.~\ref{simple}, by our observation we are also able to jointly determine the position and the velocity of the cube, without disturbing it. 

With microscopic entities, however, this is no longer possible. To understand why, we must first investigate some of the characteristics of light waves. 

Light does not exactly behave like infinitely thin and straight rays, but more like waves, and more specifically like \emph{waves of an electromagnetic nature}. Waves are characterizable by some specific parameters. In the case of the so-called \emph{plane waves}, one of these parameters is the \emph{wavelength}, usually represented by the Greek letter $\lambda$.

\begin{figure}[htbp] 
\begin{center} 
\includegraphics[width=6cm]{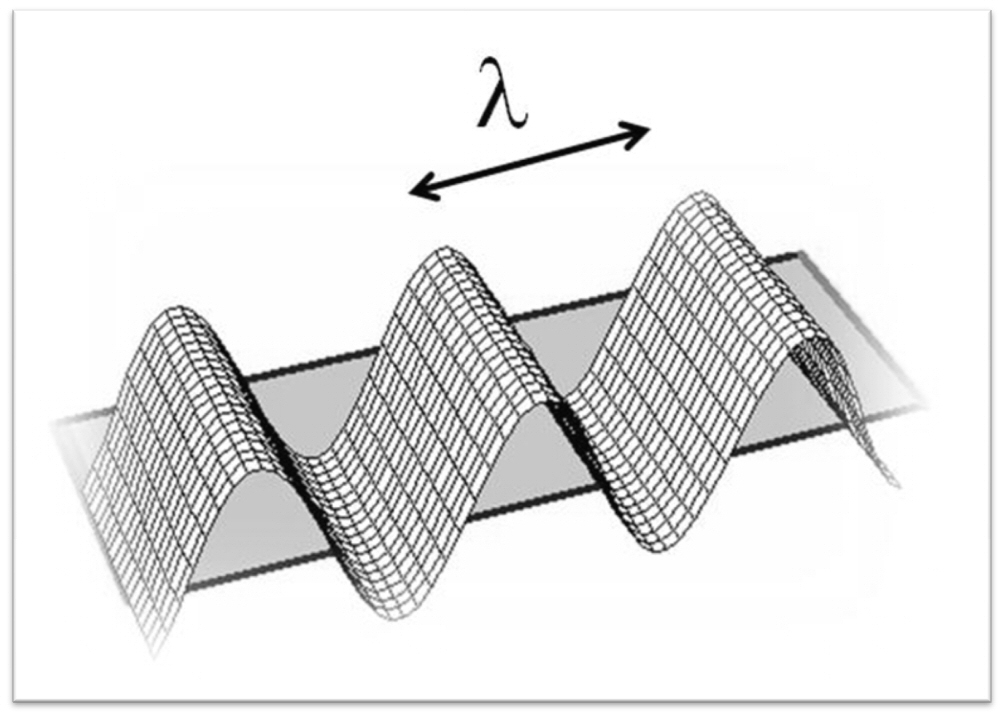} 
\caption{The wavelength $\lambda$ of a plane wave is the distance between two of its successive peaks.} 
\label{Figure6} 
\end{center} 
\vspace{-0.5cm}
\end{figure}

The wavelength $\lambda$ is nothing more than the distance between two successive peaks of the plane wave (see Figure~\ref{Figure6}). 

Let us now ask: \emph{What happens when a wave of wavelength $\lambda$ encounters an obstacle of dimension $d$?} Well, if the size $d$ of the obstacle is small compared to the wavelength $\lambda$, typically nothing will happen, in the sense that the wave will propagate undisturbed, as if the obstacle wouldn't exist. 

\begin{figure}[htbp] 
\begin{center} 
\includegraphics[width=8cm]{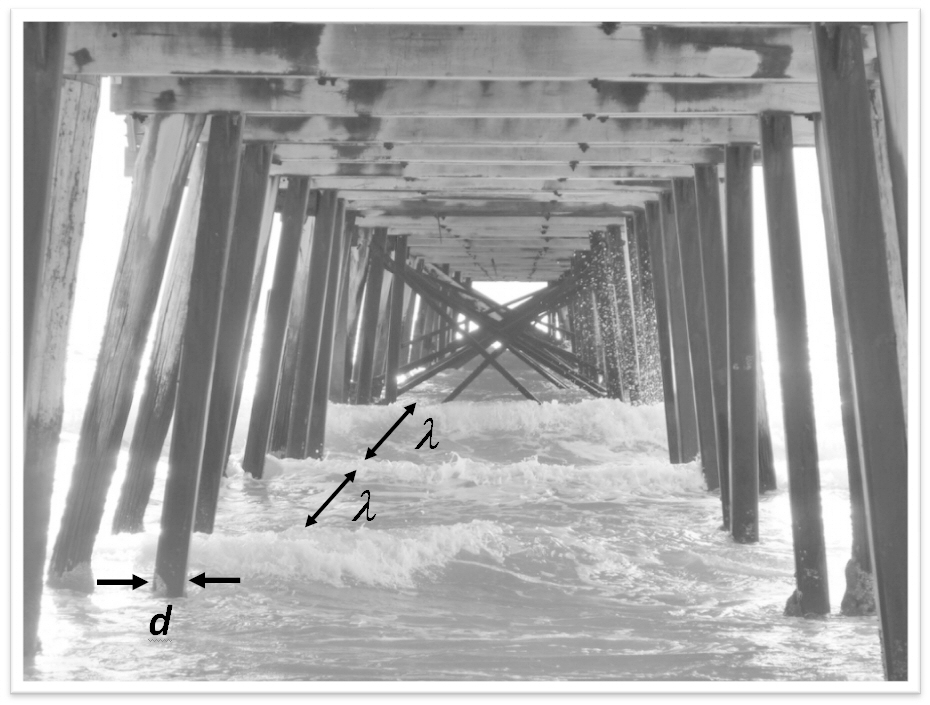} 
\caption{When the size $d$ of the obstacle is small with respect to the wavelength $\lambda$ of the wave, the latter experiences almost no deviations in its direction of propagation.} 
\label{Figure7} 
\end{center} 
\vspace{-0.5cm}
\end{figure}

Let me consider a very simple example: the waves of the sea that pass under a large pier. The poles on which rests the pier are here the obstacles. As you can observe from Figure~\ref{Figure7}, the waves propagate toward the shore totally oblivious of the poles, in the sense that in no way the poles are able to deflect their direction of propagation.

We are here in the typical situation where the obstacle's size $d$ is small compared to the wavelength $\lambda$ of the wave ($d\ll\lambda$), so that the latter cannot be detected by the former. 

What happens instead if the size $d$ of the obstacle is large compared to the wavelength $\lambda$ of the wave? To answer, consider the example of a small island. 

\begin{figure}[htbp] 
\begin{center} 
\includegraphics[width=6cm]{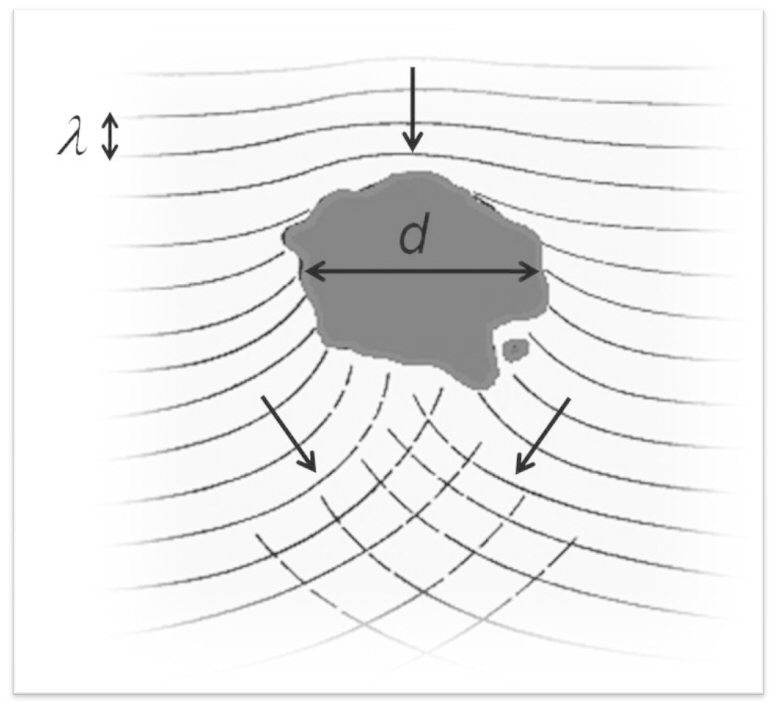} 
\caption{When the size $d$ of the obstacle is large compared to the wavelength $\lambda$ of the wave, the latter will experience important deviations in its propagation direction.} 
\label{Figure8} 
\end{center} 
\vspace{-0.5cm}
\end{figure}

As you can see on the drawing of Figure~\ref{Figure8}, which offers a top-down perspective, the wave coming from the north, winds along the two sides of the island, thus changing its direction of propagation. In this way, behind the island, a ``shadow zone" results, where the wave interferes with itself. 

We are here in the typical situation in which the obstacle's size $d$ is large compared to the wavelength $\lambda$ of the wave ($d\gg\lambda$), and because of that is able to modify in a detectable way its motion. 

Thanks to the above two examples, it should be intuitively clear now that the wavelength $\lambda$ of the wave used to see an object poses a clear limit to the precision with which it will be possible to locate it in space. In fact, if $\lambda$ is too large compared to the size $d$ of the object, the wave will not be deflected by the same, and we will have no way of knowing about its presence. This means that: 
\begin{displayquote}
\emph{The resolving power of an optical instrument can never be greater than the wavelength of the radiation used to illuminate the different objects.}
\end{displayquote}

\begin{figure}[htbp] 
\begin{center} 
\includegraphics[width=6cm]{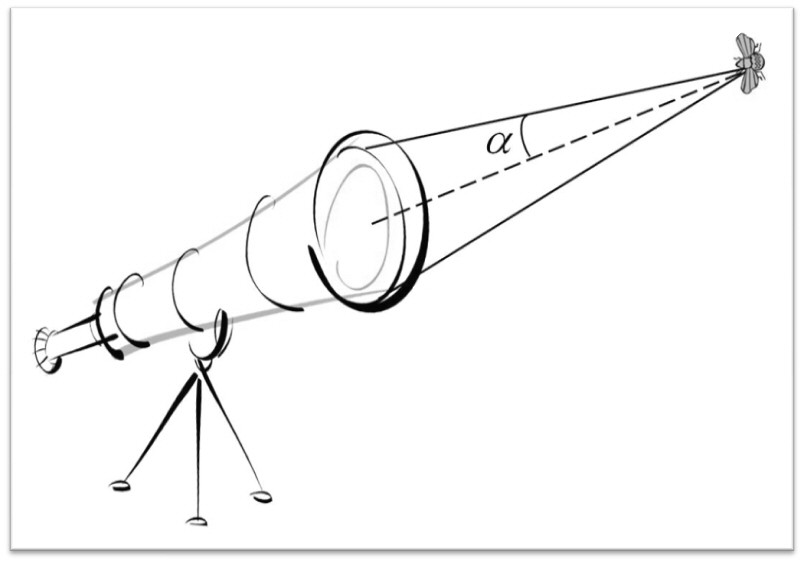} 
\caption{The opening angle $\alpha$ of an optical instrument.} 
\label{Figure9} 
\end{center} 
\vspace{-0.5cm}
\end{figure}

The \emph{resolution} of an optical instrument, however, depends not only on the wavelength, but also on the \emph{angular aperture} $\alpha$ of the instrument (see Figure~\ref{Figure9}), because of so-called and well known \emph{refractive phenomena}, which are able to blur the image of objects, thus placing a limit to the details that can be distinguished. In other words: 
\begin{displayquote}
\emph{The greater is the angular aperture $\alpha$ of an instrument and the better will be its resolution.}
\end{displayquote}

So, summing up, the resolving power of an optical instrument, such as a microscope, depends both on the wavelength $\lambda$ used and on the angular aperture $\alpha$ of the lens. We can synthesize all this with the following simple mathematical expression, stating that the minimum error $Er(x)$ we commit in determining the position $x$ of a body, because of the limited resolving power of an instrument, is directly proportional to the wavelength $\lambda$ of the radiation used, and inversely proportional to the \emph{sine}\footnote{If you have never heard of the \emph{sine function} in trigonometry, do not worry, it does not really matter for the continuation of our reasoning.} of its angular aperture $\alpha$: 
\begin{equation}
Er(x)={\lambda\over \sin\alpha}.
\label{Er(x)}
\end{equation}

\begin{figure}[htbp] 
\begin{center} 
\includegraphics[width=6cm]{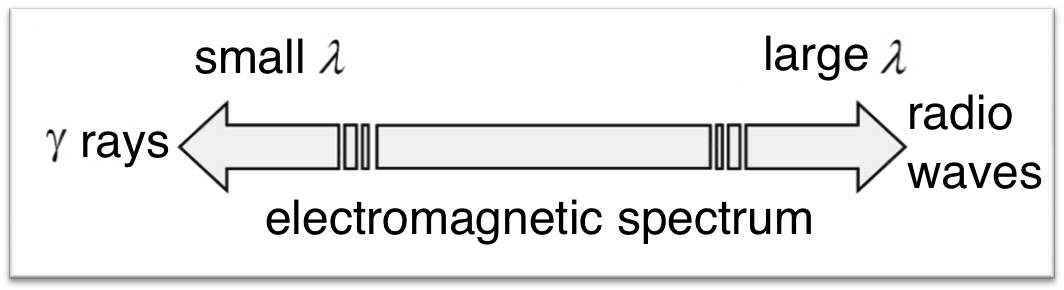} 
\caption{The electromagnetic spectrum consists of the set of all the electromagnetic radiations of different wavelengths $\lambda$.} 
\label{Figure10} 
\end{center} 
\vspace{-0.5cm}
\end{figure}

As you can see from the above simple relation, if you want to reduce the error in the determination of the position $x$, a possible strategy is evidently that of reducing the wavelength $\lambda$ of the radiation used. This is of course always possible, as an entire \emph{electromagnetic spectrum} is available, virtually infinite, ranging from radio waves of long wavelengths up to so-called gamma rays, whose wavelengths are very small (see Figure~\ref{Figure10}). 

So, using a short wavelength radiation of the gamma type, it should then be possible, at least in principle, to detect the position of the tiniest corpuscles, such as the electrons. Now, as we have seen, to see where an object is, you have to irradiate the object with an electromagnetic wave, then look at the scattered wave. In other words, the wave has to interact with the object. But what does it mean, in this specific context, to interact? 

\begin{figure}[htbp] 
\begin{center} 
\includegraphics[width=6cm]{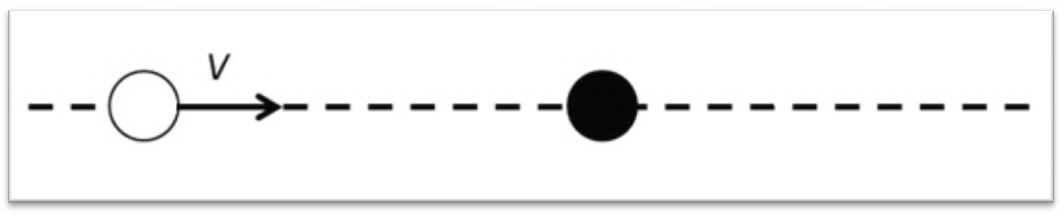} 
\caption{The situation before the collision: the white marble moves towards the black one, with speed $v$.} 
\label{Figure11} 
\end{center} 
\vspace{-0.5cm}
\end{figure}

To fix ideas, let us first consider the simple case of two marbles, able to slide on a plane without friction and without rotating. The first, of black color, is immobile, while the second, of white color, impinges the first with velocity $v$ (see Figure~\ref{Figure11}). 

\begin{figure}[htbp] 
\begin{center} 
\includegraphics[width=6cm]{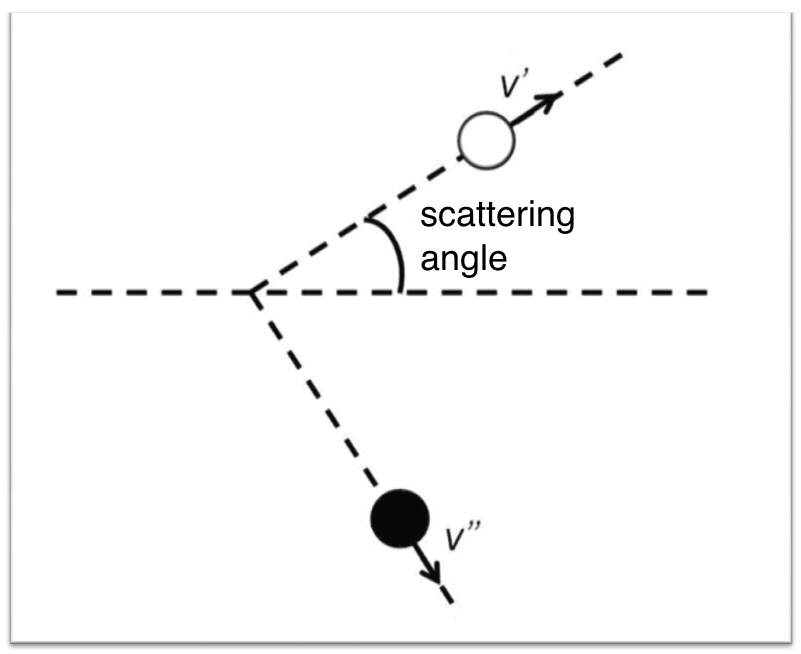} 
\caption{The situation following the collision: the white marble moves with speed $v'$, lower than the initial speed $v$, and a given scattering angle; the black marble, which received a certain amount of momentum, also moves with some speed $v''$.} 
\label{Figure12} 
\end{center} 
\vspace{-0.5cm}
\end{figure}

As you can observe in Figure~\ref{Figure12}, following the interaction, i.e., following the collision, the white marble has exchanged a certain amount of \emph{momentum}, and consequently a certain amount of \emph{energy}, with the black marble, which therefore has been set in motion, with a certain \emph{scattering angle}. 

\begin{figure}[htbp] 
\begin{center} 
\includegraphics[width=6cm]{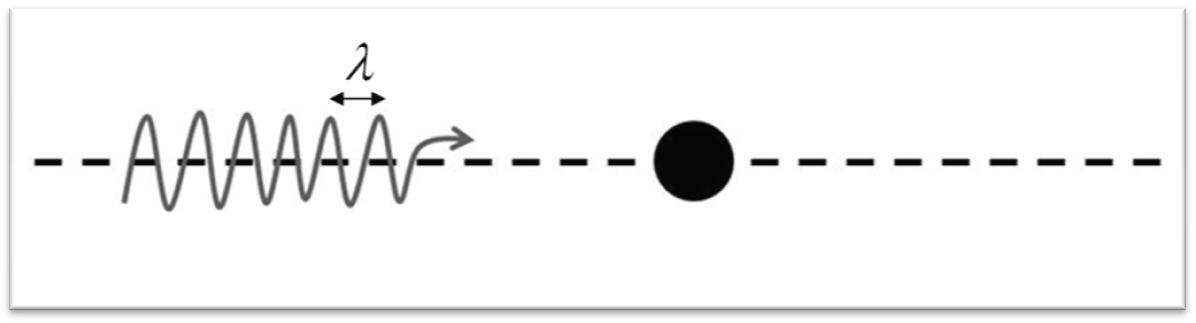} 
\caption{The situation before the collision: the electromagnetic wave (of wavelength $\lambda$) propagates (at the speed of light) in the direction of the electron.} 
\label{Figure13} 
\end{center} 
\vspace{-0.5cm}
\end{figure}

So far, everything seems clear, but what happens if the black marble, instead of being a macroscopic body, is an elementary particle, such as an electron (which for convenience I will still represent like a marble), and the white marble is not a corpuscle, but an electromagnetic wave (see Figure~\ref{Figure13})?

\begin{figure}[htbp] 
\begin{center} 
\includegraphics[width=6cm]{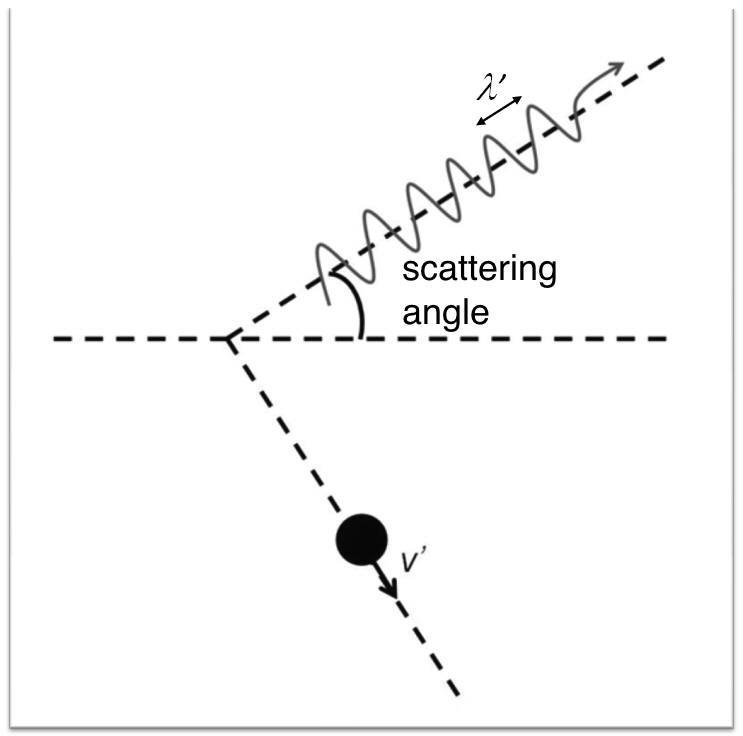} 
\caption{The situation after the collision: the electromagnetic wave is scattered with a given scattering angle and with a wavelength $\lambda'$ that is longer than the initial wavelength $\lambda$.} 
\label{Figure14}
\end{center} 
\vspace{-0.5cm}
\end{figure}

In this case, the situation after the collision is like the one represented in Figure~\ref{Figure14}. 

If you compare Figure~\ref{Figure14} with Figure~\ref{Figure12}, you will notice that the interaction process looks a lot like the previous one: the incoming wave, as if it were a marble, communicates to the electron a certain amount of momentum, also in this case by setting it in motion. 

But look more closely at what happens to the scattered wave (Figure~\ref{Figure14}): the wavelength $\lambda$ of the incident wave, following the interaction, has changed, in the sense that the wavelegth $\lambda'$ of the scattered wave is longer than $\lambda$ ($\lambda'>\lambda$). 

This effect of increase of the wavelength is called the \emph{Compton effect} (or \emph{Compton shift}), because it was discovered by the American physicist \emph{Arthur Compton} in 1923 \cite{Compton1923}. 
\begin{figure}[htbp] 
\begin{center} 
\includegraphics[width=3cm]{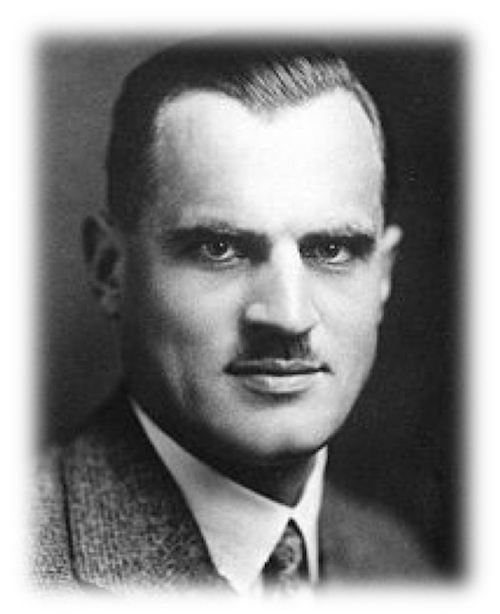} 
\caption{The American physicist Arthur Compton.} 
\label{Figure14b} 
\end{center} 
\vspace{-0.5cm}
\end{figure}

What I want here to highlight is that electromagnetic waves, similarly to moving marbles, also possess a certain amount of momentum, and when they interact with elementary particles, such as electrons, they can transfer to them part of their momentum. And when this happens, their wavelength change, in the sense that it increases.

This means, among other things, that waves possess a momentum $p$ which is inversely proportional to their wavelength $\lambda$. Translating this observation in mathematical terms, we can write:
\begin{equation}
p={h\over \lambda},
\label{momentum}
\end{equation}
where the constant of proportionality $h$ is the famous \emph{Planck constant}, whose value is really very small ($h\approx 6.626 \cdot 10^{-34}~{\rm J \cdot s}$). 

To exactly determine the value of the momentum transferred to the corpuscle, it is not sufficient, however, to just know the wavelength of the scattered wave. You also have to know the scattering angle. This is because momentum, like velocity, is a \emph{vector} quantity, and a vector quantity can vary for two distinct reasons: because its \emph{numerical value} varies, or because its \emph{direction} varies. If you do not fully understand this, no problem, it is not an essential point for understanding what follows. But what you have to keep in mind is that: \emph{to determine the momentum transferred to the corpuscle, you must also determine the scattering angle}. 

Now, the precision with which you can determine the scattering angle is limited by the angular aperture of the lens of the optical instrument you use (see Figure~\ref{Figure9}). With a simple geometry reasoning (which I leave it to the reader with some basic knowledge of trigonometric functions), it is easy to see that it is not possible to determine the amount of momentum transferred to the microscopic particle with an error $Er(p)$ lesser than the momentum $p$ of the incident wave multiplied by the sine of the diffusion angle $\alpha$. More precisely:
\begin{equation}
Er(p)\approx p\cdot\sin\alpha.
\end{equation}

As we have seen in Eq.~(\ref{momentum}), the momentum of a plane wave is simply given by the Planck constant $h$ divided by the wavelength $\lambda$. If you use this in the above expression, you obtain:
\begin{equation}
Er(p)\approx {h\over \lambda}\cdot\sin\alpha.
\end{equation}

At this point, if you remember that the momentum of the particle is given by the product of its mass $m$ times its velocity $v$ ($p=m\cdot v$), you can divide by $m$ on the right and left sides of the above expression, thereby obtaining an estimate of the minimum error $Er(v)$ on the velocity:
\begin{equation}
Er(v)={Er(p)\over m}\approx {h\over \lambda\cdot m}\cdot\sin\alpha.
\label{Er(v)}
\end{equation}

On the other hand, considering the previously derived expression (\ref{Er(x)}) for the minimum error $Er(x)$ on the position of the particle, multiplying it by (\ref{Er(v)}), you get: 
\begin{equation}
Er(v)\cdot Er(x)\approx {h\over \lambda\cdot m}\cdot\sin\alpha \cdot {\lambda\over \sin\alpha} = {h\over m}.
\end{equation}

This is nothing but (\ref{HUP}), with the constant $c={h\over m}$ (not to be confused with the speed of light) whose value for an electron is about 1 cm$^2$/s. 
In other words, you have just derived the famous HUP.

\section{Complementarity and incompatibility}

\begin{figure}[htbp] 
\begin{center} 
\includegraphics[width=3cm]{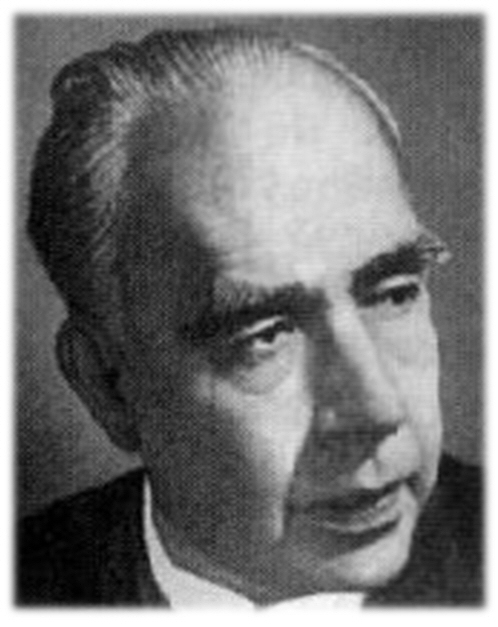} 
\caption{The Danish physicist Niels Bohr.} 
\label{Figure15} 
\end{center} 
\vspace{-0.5cm}
\end{figure}

\noindent The strange situation expressed by the HUP (and many other situations that are encountered when one studies microscopic systems) has been summarized in 1928 by the Danish physicist \emph{Niels Bohr} (see Figure~\ref{Figure15}) in his famous \emph{principle of complementarity} \cite{Bohr1928}. 

Roughly speaking, this principle states that: 
\begin{displayquote}
\emph{There are properties that are mutually exclusive, and therefore cannot be observed simultaneously, using a same experimental arrangement, i.e., within the same experimental context}. 
\end{displayquote}

So, if we measure (that is, if we observe, in a practical way) with good precision the position property of a particle, we will automatically and inevitably alter in a profound and totally unpredictable way (or even make it indeterminate) its velocity property, and vice versa. The two properties -- possessing a given position and possessing a given speed -- being in a sense complementary, they cannot be jointly observed.

At this point, two remarks are in order. The first is that this possible alteration, for example of the velocity when we observe the position, takes place in a completely unpredictable way, that is, in a way that is not determinable a priori by the observer. 
This aspect of the unpredictability does not emerge directly from our simplified analysis of the Compton Effect. However, it is an integral part of the formalism of quantum mechanics. 

To put it simply, according to quantum theory: \emph{we cannot determine in advance what will be, for example, the scattering angle following the collision, and can only calculate the probabilities associated with the different possible scattering angles}. 

The second remark is that in all our reasonings we have implicitly assumed that the microscopic particle always possessed, even before being observed, a specific position and a specific velocity, although these were not known by the observer/experimenter. As we are going to see, this assumption is however completely unfounded. 

Having said that, let us try to understand the concept of complementarity a little better. The word ``complementarity" is obviously quite attractive from a philosophical point of view, and in part is certainly correct, but it can also lead to a possible misrepresentation of the issue we are analyzing here. Instead of the term ``complementarity," you can use the simpler and more direct term of ``incompatibility," to be understood in the sense of the incompatibility of the procedures of observation of certain properties, associated with specific experimental arrangements.

In physics, one tries to make precise the concept of experimental \emph{incompatibility} by using the idea of \emph{non-commutability}. More precisely: 
\begin{displayquote}
\emph{If two observations are compatible, the order with which you perform them does not affect their outcomes (and therefore such order can be freely switched). When instead a change in the order of the observations can affect their outcomes, the two observations are said to be incompatible}.
\end{displayquote}
This is exactly what happens with the position and the velocity of a microscopic particle: to observe first the position and then the velocity does not produce the same results than to observe (i.e., to measure) first the velocity and then the position. This is mainly due to the fact that these observations are invasive processes, modifying the state of the observed entity in an unpredictable way. 

It is however important to understand that the incompatibility I'm here talking about is not a feature of the microscopic processes only: it can also manifest in many of the operations we perform every day. Let me consider a simple example. 

\begin{figure}[htbp] 
\begin{center} 
\includegraphics[width=6cm]{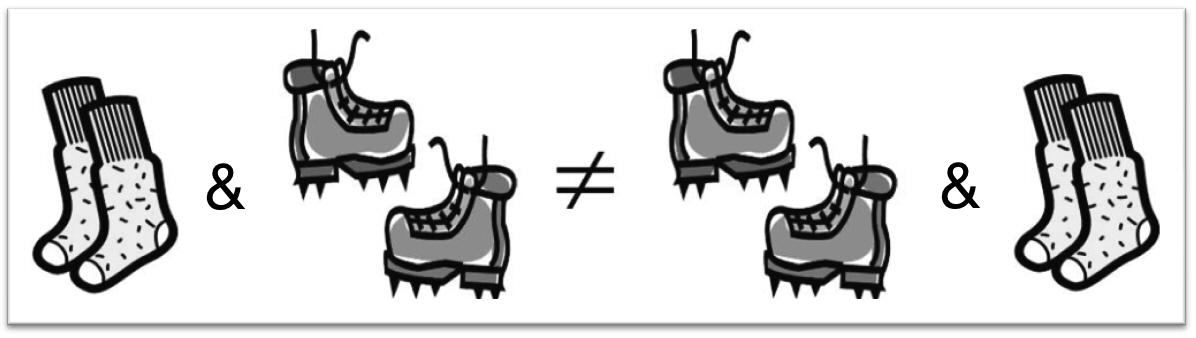} 
\caption{The operations ``putting on the socks'' and ``putting on the shoes" are non-commutative (the symbol ``$\neq$'' means ``not equal'').} 
\label{Figure16} 
\end{center} 
\vspace{-0.5cm}
\end{figure}

I hope you will agree that to put on the socks first, then the shoes, does not produce the same outcome as to put on the shoes first, then socks (see Figure~\ref{Figure16}). 

These two processes being non-commutative (their order of execution is crucial for the final result), they can be considered to be mutually incompatible. But of course, not all processes are mutually incompatible. Many are perfectly compatible. Let me consider a simple example of two perfectly compatible operations, that is, two operations whose order of execution can be switched, without affecting the final result. 

I hope you will agree that to put on the socks first, then the gloves, is the same, i.e., it produces the same result, than to put on the gloves first, then the socks. 
\begin{figure}[htbp] 
\begin{center} 
\includegraphics[width=6cm]{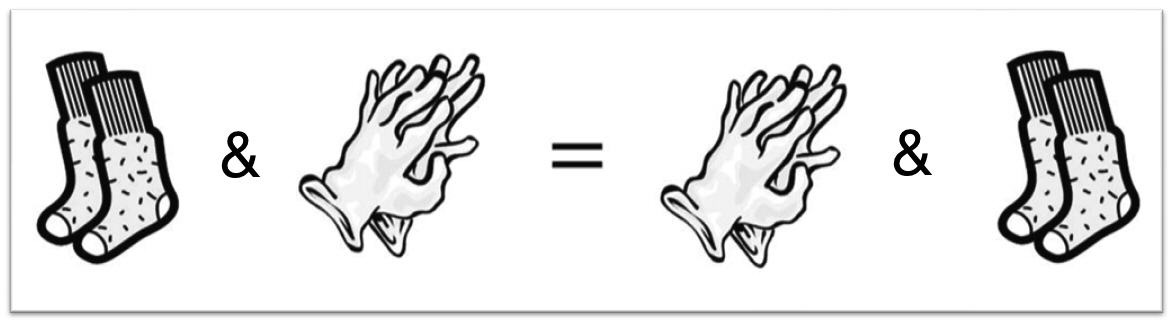} 
\caption{The operations ``putting on the socks'' and ``putting on the gloves" are commutative.} 
\label{Figure17} 
\end{center} 
\vspace{-0.5cm}
\end{figure}

It is instructive to consider some further examples of operations that are mutually incompatibles. A novice asked the prior: ``Father, can I smoke when I pray?" And he was severely reprimanded. A second novice asked the prior: ``Father, can I pray when I smoke?" And he was praised for his devotion. 

\begin{figure}[htbp] 
\begin{center} 
\includegraphics[width=6cm]{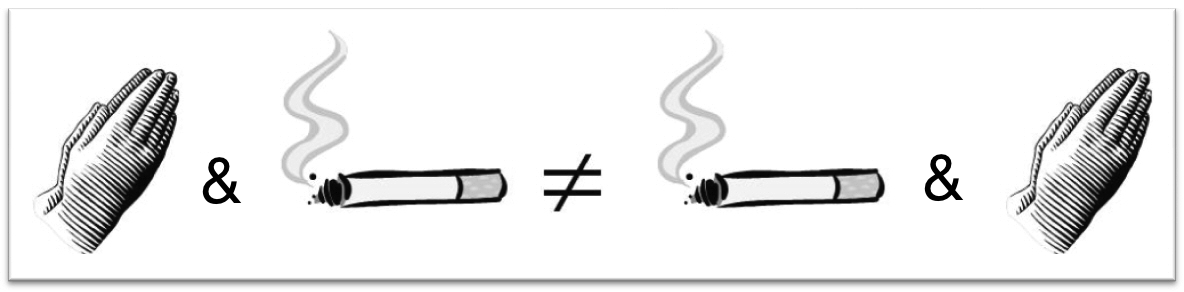} 
\caption{The operations ``to pray'' and ``to smoke" are non-commutative, according to the prior's understanding.} 
\label{Figure18} 
\end{center} 
\vspace{-0.5cm}
\end{figure}

In other words, to pray and smoke does not produce the same effect as to smoke and pray (see Figure~\ref{Figure18}). Here the non-commutability is expressed through the order chosen for the verbs ``to smoke" and ``to pray" in a sentence. If you switch the order of the verbs, it also changes the perceived sense of the phrase. This because verbs indicate actions, that is, operations that we perform, which is the reason why their order in a sentence is often so crucial. 

What we must understand is that, in general, the order with which we operate in reality affects the final outcome. To assemble an IKEA piece of furniture, it is necessary to operate in exactly the sequence indicated in the instruction manual, if you want to obtain the desired result. 

\begin{figure}[htbp] 
\begin{center} 
\includegraphics[width=6cm]{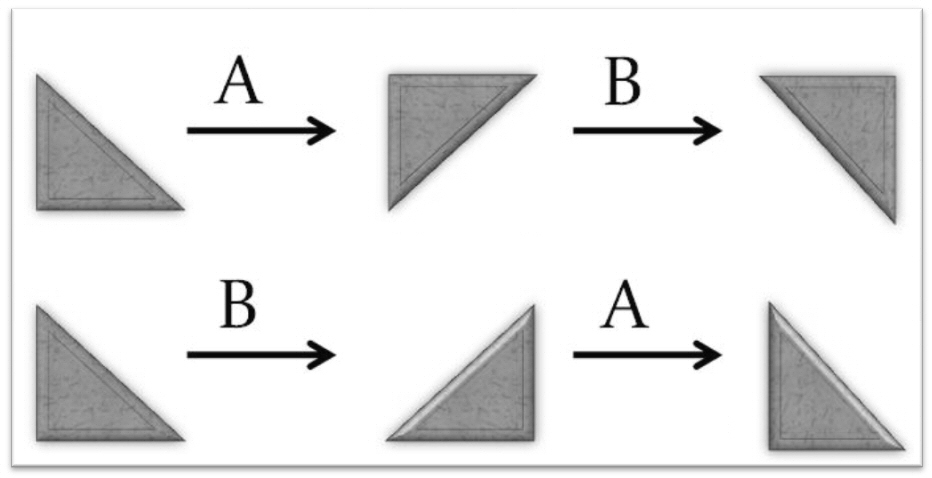} 
\caption{The ``rotation" and ``reflection" operations are non-commutative, as depending on their order they will in general not produce the same result.} 
\label{Figure19} 
\end{center} 
\vspace{-0.5cm}
\end{figure}

To make sure that this issue is fully understood, let me consider still another example, using a simple \emph{right triangle} (a triangle in which one angle is $90^\circ$). I define the operation $A$ as consisting in rotating the triangle $90^\circ$ clockwise.
The operation $B$, instead, is by definition a reflection of the figure with respect to the vertical axis. As you can see in Figure~\ref{Figure19}, depending on the order of the operations, the final result will not be the same: $A$ and $B$ are therefore incompatible operations, being non-commutative operations. 

\section{Measuring a wooden cube}

Having clarified the concept of incompatibility, and the fact that incompatibility can be expressed in terms of non-commutability, let me show what are the consequences of all this when you try to observe two specific properties of an ordinary macroscopic entity, to you now familiar: a wooden cube \cite{Aerts1982}.

So, the physical system you are about to study/observe is a simple wooden cube. With the letter $A$, you decide to denote the process of observation of the property of the cube of \emph{burning well}. On the other hand, with the letter $B$, you also denote the process of observation of the property of the cube of \emph{floating on water}. 

There are of course different possible ways to define these two properties of burning well and floating on water. So, if we want to be more precise, we must specify what you mean in practice, for the cube, to have these two properties tested, i.e., what are the operations you have to exactly perform, and the results you have to obtain, to successfully observe these two properties.

\begin{figure}[htbp] 
\begin{center} 
\includegraphics[width=6cm]{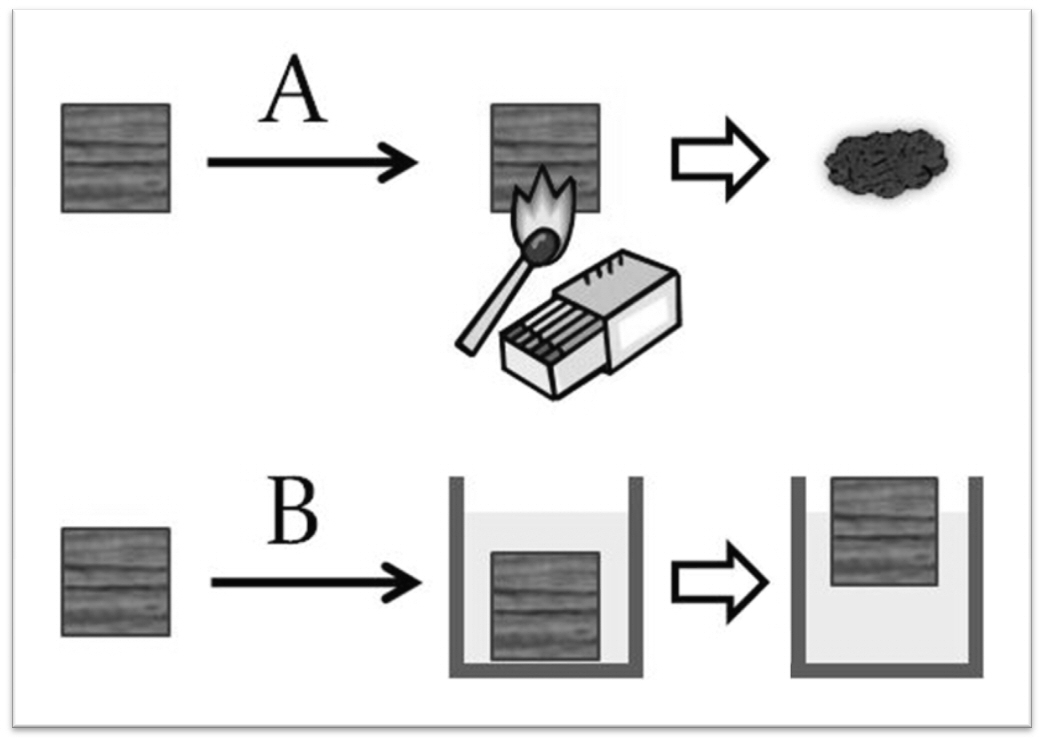} 
\caption{Operations $A$ and $B$, and their respective outcomes, when carried out on a wooden cube.} 
\label{Figure20} 
\end{center} 
\vspace{-0.5cm}
\end{figure}

For example, you can decide that the observation of the burning well property involves exposing the cube to the flame of a match, for a few seconds. If, following this operation, the cube is set on fire and reduces to ashes, the observation of the burning well property is considered to be successful, and you can say that the property has been confirmed (see Figure~\ref{Figure20}).

You can also decide that the observation of the floating on water property consists in completely immersing the cube in a container filled with water and then check if thanks to Archimedes' \emph{buoyant force} it raises to the surface. If this happens, the observation of the floating property is considered to be successful, and you can say that the property has been confirmed (see Figure~\ref{Figure20}). 

Considering our previous discussion, it is then natural to ask the following question: Does the wooden cube possess \emph{both} properties: burning well and floating on water? (Which for reasons of conciseness, we will also call \emph{burnability} and \emph{floatability}, respectively).

\begin{figure}[htbp] 
\begin{center} 
\includegraphics[width=10cm]{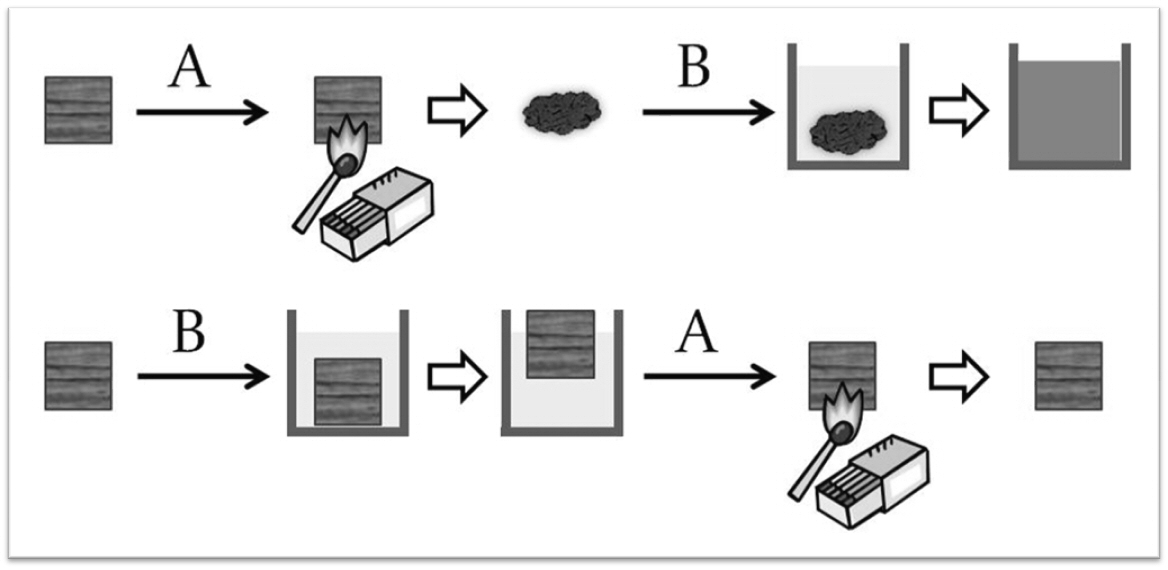} 
\caption{Operations $A$ and $B$, depending on their order of execution, produce different outcomes. If $A$ is successful, then the outcome of $B$ will be negative and. vice versa, if $B$ is successful, then the outcome of $A$ will be negative.} 
\label{Figure21} 
\end{center} 
\vspace{-0.5cm}
\end{figure}

One way to possibly verify this consists in taking the wooden cube and then try to observe these two properties, one after the other. If you start with burnability, you can see that the cube burns well, i.e., that it turns into a small pile of ashes. But then, if you try to observe its floatability, plunging the obtained ashes into the water, these will not raise to the surface, since, as is well known, ashes do not float (see Figure~\ref{Figure21}). So, the conclusion of the above sequence of observations is that the wooden cube burns well, but does not float. 

You could then try reversing the order of the two observations. If you start with the floatability, you can see that the wooden cube floats easily. However, if you subsequently want to observe its burnability property, subjecting it to the flame of the match, it will not burn, since a wet cube, as is known, does not burn well (see Figure~\ref{Figure21}). So, the conclusion of this second sequence of observations is that the wooden cube floats on water, but does not burn well. 

What you have just pointed out is the simple fact that the observational processes $A$ and $B$, associated with the burnability and floatability properties, respectively, do not commute. In other words, they are mutually incompatible. Therefore, it does not seem possible to jointly observe the burnability and the floatability of the wooden cube. 

If this is true, as it is true, then you may wonder: Does the wooden cube really possess, at once, both properties of burnability and floatability? 

The question is legitimate, since apparently you are unable to jointly observe these two different properties. On the other hand, according to your intuition, the wooden cube certainly possesses at once both properties of burning well and floating on water. In the same way for example a car can be at the same time crash-proof and 4 meters long. More precisely: 
\begin{displayquote}
\emph{Intuition tells you that an entity can possess at once a number of different properties, although not all of them are necessarily observable at the same time, or one after the other}.
\end{displayquote}

Very well, let us recap. It is clear that the cube possesses the property of burning well. It possesses such property because if you execute the test $A$ that by definition allows to observe it, the test will invariably be successful, so the property will be confirmed. In the same way, it is clear that the cube possesses the property of floating on water. It possesses such property because if you execute the test $B$ that by definition allows to observe it, the test will invariably be successful, so the property will be confirmed. 

\begin{figure}[htbp] 
\begin{center} 
\includegraphics[width=3.5cm]{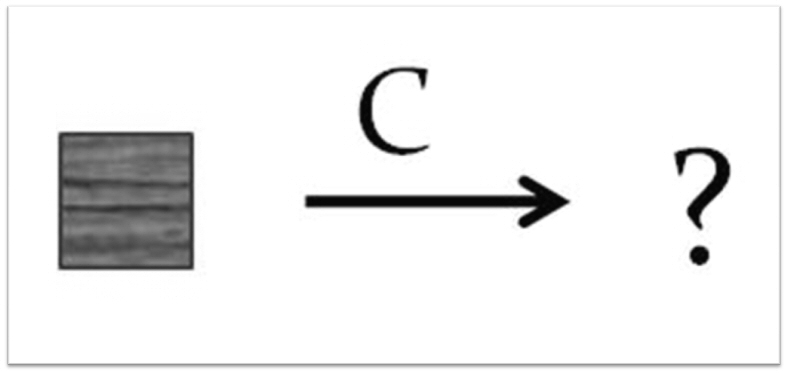} 
\caption{The mysterious operation $C$, allowing to test the meet property of burning well and floating on water.} 
\label{Figure22} 
\end{center} 
\vspace{-0.5cm}
\end{figure}

Perfect, but since your intuition also tells you that the cube possesses the \emph{meet property} of burning well \emph{and} floating on water, it is natural to ask: What would be the test $C$ allowing to confirm the meet property of burning well and floating on water? (See Figure~\ref{Figure22}). In other words: How can you know if your intuition is correct and that it is true that the cube possesses at once these two properties, despite the fact that they are mutually incompatible? 

\begin{figure}[htbp] 
\begin{center} 
\includegraphics[width=6cm]{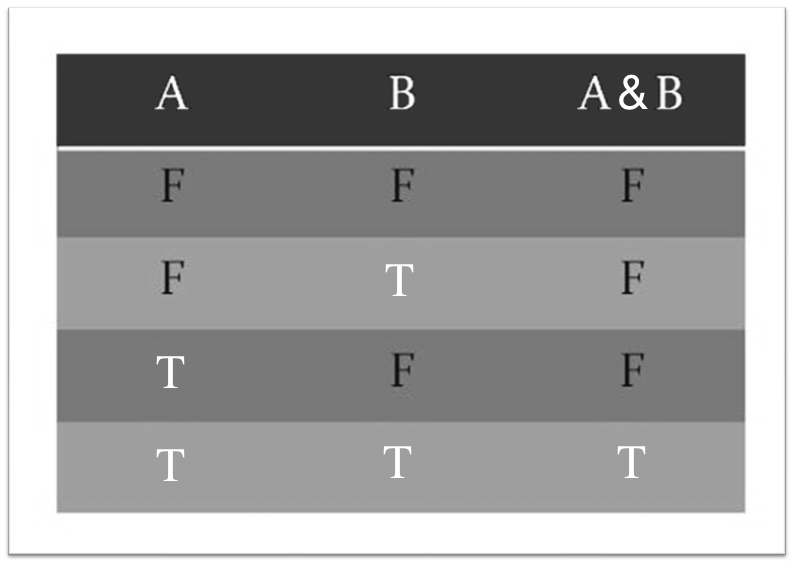} 
\caption{The truth table of classical logic, here for the conjunction logical operator ``and.''} 
\label{Figure23} 
\end{center} 
\vspace{-0.5cm}
\end{figure}

Apparently, you are confronted here with a little puzzle. Indeed, if you take a look at the \emph{truth tables} of \emph{classical logic}, and more particularly the truth table for the \emph{conjunction}, described in Figure~\ref{Figure23}, you can observe the following. Reading the first line of the table, you find that if a property $A$ is false, and a property $B$ is false, then, inevitably, also the meet property $A$ \emph{and} $B$" is false. In other words, in classical logic the meeting of two falsities is once again a falsity. 

But, as evidenced by the second and third line of the table, it is also sufficient that only one of the two properties is false, to make the associated meet property, as a whole, false. In fact, as made evident in the last line of the table, the meet property $A$ \emph{and} $B$ can be true if and only if both property $A$ and $B$ are individually true. 

On this obviously you can only agree. But considering that burnability and floatability are mutually incompatible properties, how can you bring out their joint trueness. In other words: How can you test together, conjunctly, at once, the truth value of two properties that are experimentally mutually incompatibles?

I hope it is clear what is the relevance of this discussion in relation to the previous analysis of Heisenberg uncertainty principle (HUP), on which of course I will be back shortly. In fact, if you remember, the position and velocity of a microscopic particle are linked by an uncertainty relation; a relation which in turn expresses a condition of incompatibility. 

Does this mean that a microscopic particle is not able to jointly possess a position and a velocity? In other words: Is HUP a statement about the simultaneous non-existence of the position and velocity of a microscopic entity, such as an electron, or is it just a statement about our limitation in jointly knowing these two physical quantities?

Based on your intuition about the burnability and floatability of a wooden cube, I'm sure you would be tempted to say that the experimental incompatibility of two physical quantities does not mean that they cannot exist simultaneously, and that therefore nothing prohibits an electron to simultaneously possess a well-defined position and velocity. 

Is the above correct? To find out, I propose to continue our conceptual analysis, and for this it will be useful to define a bit more clearly what a \emph{property} is.

\section{The EPR-PA reality criterion}

In general, you can say that a property is something that an entity can have, which can be observed, and is defined, by means of an \emph{experimental test}, whose execution enable one to confirm (or to invalidate) the property in question. But be careful: 
\begin{displayquote}
\emph{To confirm a property does not necessarily mean to prove that the property is or was actual, in the sense that the property is or was stably possessed by the entity in question.} 
\end{displayquote}

\begin{figure}[htbp] 
\begin{center} 
\includegraphics[width=9cm]{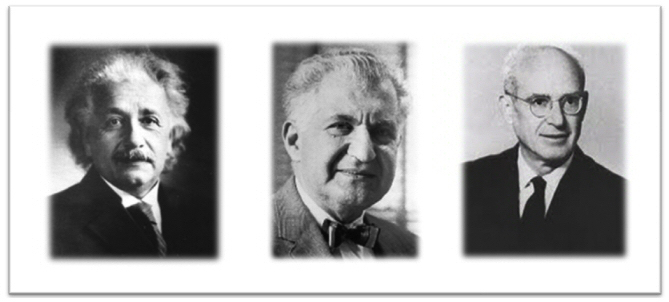} 
\caption{From left to right, physicists Albert Einstein, Boris Podolsky and Nathan Rosen.} 
\label{Figure24}
\end{center} 
\vspace{-0.5cm}
\end{figure}

As the wise would say: a burnt wooden cube is no longer a burnable wooden cube! That said, to continue your exploration, you now need the help of the German physicist \emph{Albert Einstein} and of his two Russian and American-Israeli collaborators, \emph{Boris Podolsky} and \emph{Nathan Rosen} (see Figure~\ref{Figure24}). 

In a famous article published in 1935 \cite{Einstein1935}, these three scientists enunciated an important \emph{reality criterion}, which is of course also an \emph{existence criterion}, since in the common understanding of these two concepts, something is considered to be real if and only if it exists. 

More precisely, Einstein, Podolsky and Rosen (in brief, EPR), in this their famous article, said the following:
\begin{displayquote}
``If, without in any way disturbing a system, we can predict with certainty [...] the value of a physical quantity, then there exist an element of physical reality corresponding to this physical quantity.''
\end{displayquote}

What EPR have clearly recognized is that our description of reality is essentially based on our reliable predictions about it. However, they also remained rather cautious regarding their criterion, as they added in their article that: 
\begin{displayquote}
``It seems to us that this criterion, while far from exhausting all possible ways of recognizing a physical reality, at least provides us with one such way, whenever the conditions set down in it occur. Regarded not as a necessary, but merely as a sufficient, condition for reality, this criterion is in agreement with classical as well as quantum-mechanical ideas of reality.''
\end{displayquote}

Despite their warning, EPR did not offer a single counter example of what would be the nature of an \emph{element of physical reality} not subject to their criterion. In other words, although they assumed, very prudently, that their criterion was only sufficient, they presented no reasons as to why it should not be considered, at least in principle, also necessary. 

But let me come back once more to the wordings of the criterion. An important point to emphasize is that when EPR write ``if [...] we can predict with certainty," what one should understand is: ``if we can \emph{in principle} predict with certainty." Indeed, the important point is not if one possesses in practice all the information allowing to make a reliable prediction, but if this information is available somewhere in the universe (although maybe dispersed who know where), so that a being of sufficient power and intelligence could in principle access it. 

\begin{figure}[htbp] 
\begin{center} 
\includegraphics[width=9cm]{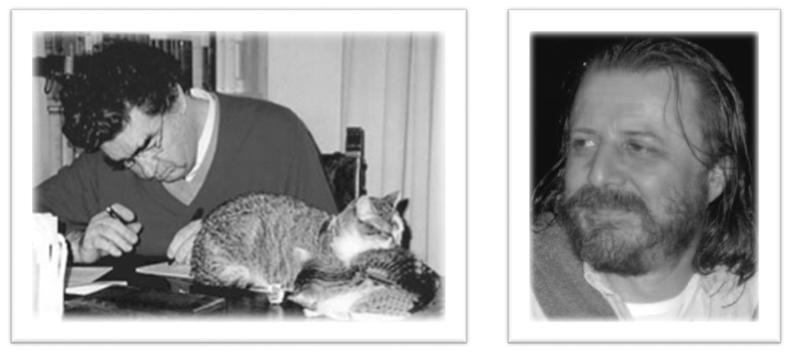} 
\caption{Belgian physicists Constantin Piron (left) and Diederik Aerts (right).} 
\label{Figure25}
\end{center} 
\vspace{-0.5cm}
\end{figure}

That said, it is worth observing that this important criterion of reality, or of existence, was subsequently reconsidered by the Belgian physicists \emph{Constantin Piron} \cite{Piron1976} and \emph{Diederik Aerts} \cite{Aerts1982} (see Figure~\ref{Figure25}), the latter being also the author of the paradigmatic example of the floatability and burnability of a wooden cube. 

These two physicists reformulated the EPR criterion in a much more specific and complete form, which is roughly the following \cite{Aerts1982,Piron1998,Sassoli2011}: 
\begin{displayquote}
\emph{If, without in any way disturbing the physical entity under consideration, it is in principle possible to predict with certainty the successful outcome of an experimental test, then the property associated with that test is an actual (existing) property of the physical entity. And vice versa}.
\end{displayquote}

In other words, if the property of a physical entity is \emph{actual}, then (without in any way disturbing the entity) it is in principle possible to predict with certainty the successful outcome of an observational test associated with it. Therefore, according to this more complete reality criterion, which I will simply call the \emph{Einstein-Podolsky-Rosen-Piron-Aerts (EPR-PA) criterion}: a property is actual if and only if, should one decide to perform the observational test that defines it, the expected result would be certain in advance. 

This means that the entity has the property in question before the test is done, and in fact even before one would have chosen to execute it, which is the reason why one is allowed to say that the property is an \emph{element of reality}, existing independently from our observation. 

On the other hand, if one cannot apply the EPR-PA criterion, i.e., if one cannot, \emph{not even in principle}, predict the outcome of the test defining the property in question, one must conclude that the entity under consideration does not possess that property, i.e., that the property is not an actual (existing) one, but only a \emph{potential} property (if the probability of actualizing the property, in some experimental contexts, is non-zero). 

Again, the above conclusion is correct provided the prediction cannot be made even in principle. Indeed, in most experimental situations one simply do not possess a complete knowledge of the entity, and therefore one does not have access to all its actual properties. But when one possesses a complete knowledge of the entity, then by definition one is also able to predict with certainty all that is predictable about it, so that what cannot be predicted is, by definition, a non-existing (potential, uncreated) aspect of reality.

After this important detour on the issue of reality criteria, it is time to return to our little wooden cube. Based on the EPR-PA criterion, to determine whether the wooden cube possesses or not the property of burning well, you do not have to execute the burnability test, and burn it, but simply be in a position to predict with certainty that, should you perform the test, the outcome would be certainly positive.

\begin{figure}[htbp] 
\begin{center} 
\includegraphics[width=6cm]{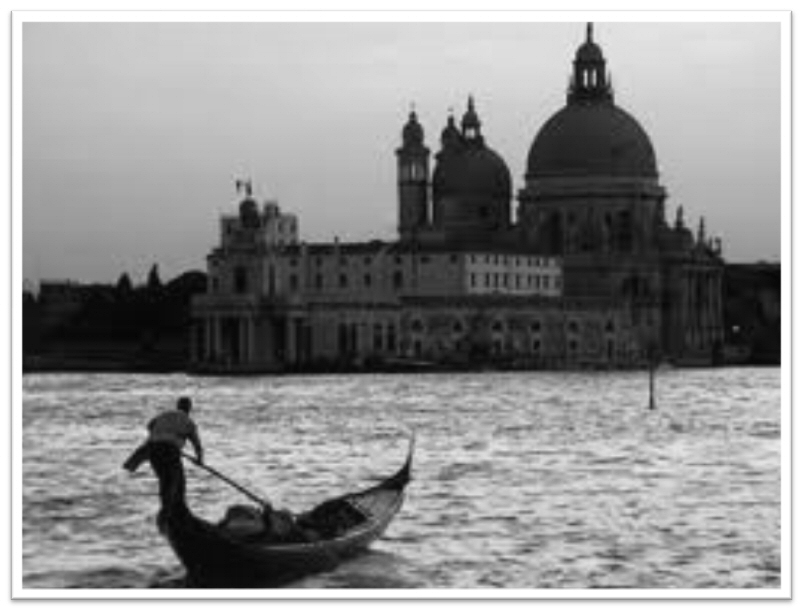} 
\caption{A typical postcard image of Venice (Italy).} 
\label{Figure26}
\end{center} 
\vspace{-0.5cm}
\end{figure}

Consider a much more straightforward example: think about the city of \emph{Venice} in \emph{Italy} (see Figure~\ref{Figure26}), now, and ask yourself: Does Venice exist right now? I'm assuming that in this moment you are not in Venice, that is, that you are not having an experience with Venice in this moment.

In other words, you cannot base your claim about the existence of Venice now, on the fact that you would be having, now, an experience with it. But this is not necessary, as per the EPR-PA criterion, to be able to affirm that Venice exists in this moment, all you need is to be able to predict that, should you have decided to have an experience with Venice now (for example by organizing a trip a few days ago, to go to Venice, so that you could be there now), the latter, with certainty, would have been available to be part of it. So, summing up: 
\begin{displayquote}
\emph{Reality (what exists) does not only correspond to the phenomena that you factually experience. Reality also, and above all, corresponds to all the possible phenomena: those that you could have in principle experienced with certainty, if only you would have chosen to do so in your past.} 
\end{displayquote}

Reality, therefore, is constructed in a \emph{counterfactual} way. In the sense that you can speak in a perfectly meaningful way even of things you are not currently concretely observing, provided that, if you would have decided to do so, the positive result of the observational process would have been absolutely certain in advance. 

How can you apply all this to the problem of our wooden cube? More particularly, how can you use this to solve the problem of demonstrating that the wooden cube actually possesses the meet property of burning well and floating on water, although these two properties are individually experimentally mutually incompatible? 

According to the EPR-PA criterion, it is sufficient to be able to predict that, should you perform the test $C$ associated with such meet property, the positive outcome would be certain in advance. All right, but what would be then this mysterious test $C$, associated with the meet property of burning well and floating on water? In other words: What is, generally speaking, the observational test of a meet property ``$A$ and $B$"?

The specifications of this particular test, which in technical language is called a \emph{product test}, were given some years ago by Constantin Piron \cite{Piron1976,Piron1998}, who I mentioned in relation to EPR's reality criterion. Let me explain what a product test is. 

It is very simple: first of all, you need an instrument that can generate, in a completely \emph{random} way, two events, which I will simply call ``heads'' and ``tails.'' For example, the instrument could be the toss of a coin, provided it is carried out in such a way as not to allow you (the experimenter) to predict the outcome in any way. 

\begin{figure}[htbp] 
\begin{center} 
\includegraphics[width=8.5cm]{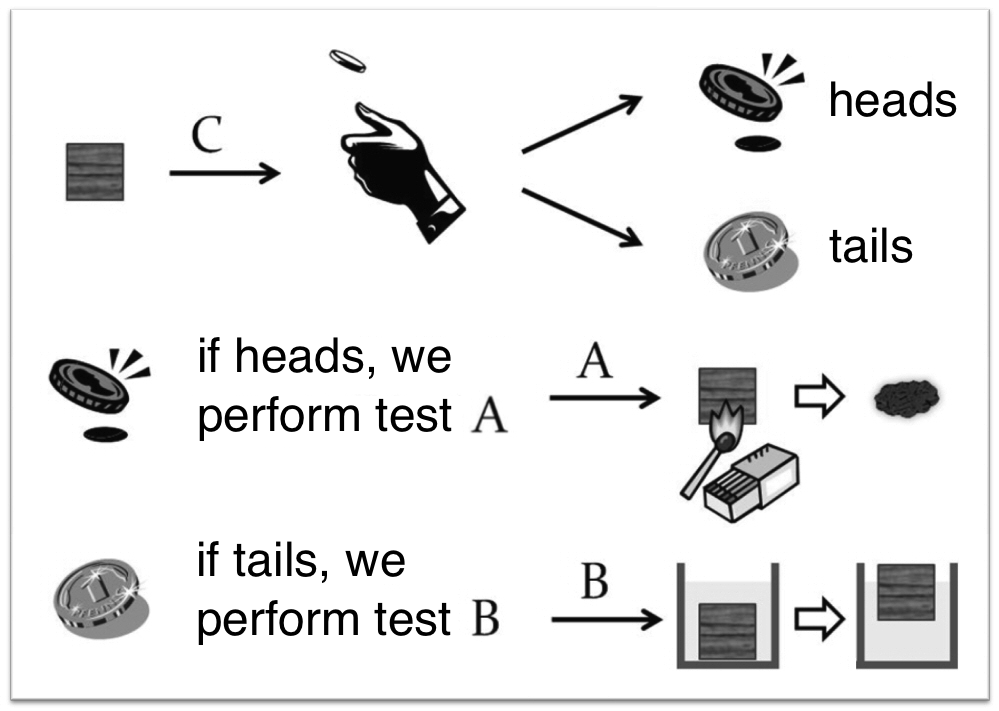} 
\caption{Graphical representation of the execution logic of a product test, able to test the meet property of burning well \emph{and} floating on water.} 
\label{Figure27}
\end{center} 
\vspace{-0.5cm}
\end{figure}

So, if following the toss of the coin you obtain ``heads,'' you perform test $A$, and the outcome (positive or negative) will be assigned to the test $C$, of the meet property. On the other hand, if you obtain ``tails,'' you perform test $B$, and again the outcome (positive or negative), will be attributed to $C$ (see Figure~\ref{Figure27}). 

Here you are, now you know the specific nature and logic of the observational test of a meet property, called a product test. Perhaps you will object that, since following the coin toss you only have to perform one of the two tests, this means that only one of the two properties would in fact be tested. This is not exact, as you do not have to forget that the toss of the coin is an integral part of the observational process, and that the choice of which of the two tests will be performed is totally unpredictable. 

Therefore, the only way to ensure a priori, with certainty, the positive outcome of test $C$ (without the need of executing it), is to have the cube possessing both properties at once, that is, possessing them at the same time!

So, when you go from properties to tests, the conjunction logical operator ``and" transforms into the disjunction logical operator ``or:" to test the meet property $A$ \emph{and} $B$, you have to test either $A$ \emph{or} $B$, choosing however one of these two alternatives -- and this is really the crucial point -- totally at random. 

That said, what can you say then about the meet property of the cube of jointly burning well and floating on water? Does it possess it, in actual terms, or not? 
Evidently, according to the test product $C$ just defined, and to EPR-PA reality criterion, as you are able to predict with certainty the positive outcome of both tests, of burnability and floatability, you can deduce that the wooden cube jointly possesses these two properties, at once, although they are mutually experimentally incompatible. 

In summary:
\begin{displayquote}
\emph{The experimental incompatibility of two properties $A$ and $B$ of a given entity does not necessarily imply that they cannot be simultaneously actual, as the disjunctive logic of a product test shows and the example of the wooden cube demonstrates.} 
\end{displayquote}

As a result, you may now be tempted to conclude that, even though the position and velocity of a microscopic corpuscle are incompatible physical quantities, as expressed by HUP, nevertheless, it should be possible to consider them to be simultaneously actual. But such consideration would be completely wrong.

\section{Non-spatiality}

Let me try to explain why the position and velocity of a microscopic particle, contrary to the burnability and floatability of a wooden cube, cannot be considered properties that are jointly possessed by a microscopic entity like an electron. For this, let me start considering again the case of a macroscopic body, that is, of a body of large dimensions, visible with the naked eye. 

\begin{figure}[htbp] 
\begin{center} 
\includegraphics[width=6cm]{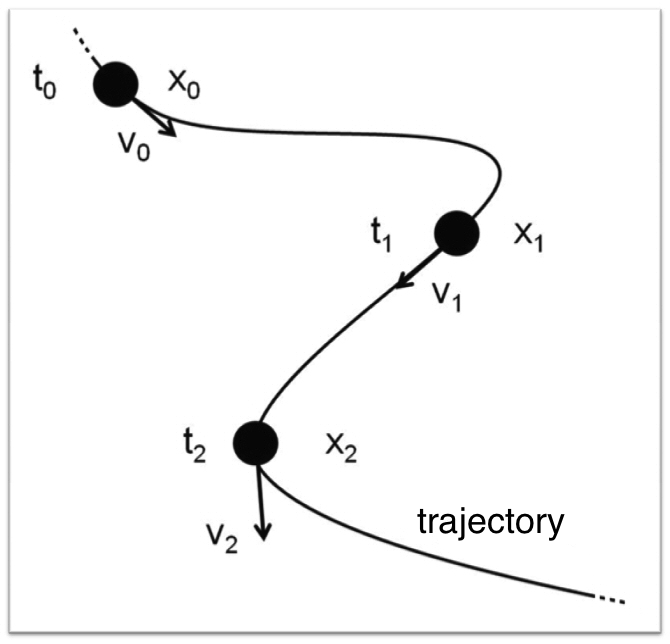} 
\caption{A graphical representation of the \emph{trajectory} traveled in space by a moving macroscopic body, i.e., of the different positions taken by the body over time.} 
\label{Figure28}
\end{center} 
\vspace{-0.5cm}
\end{figure}

At time $t_0$, the body is, say, located in a position $x_0$ and, at that same instant, it also possesses a well-defined velocity $v_0$ (see Figure~\ref{Figure28}). 

As previously discussed, you know that when you both know the position ($x_0$) and the velocity ($v_0$) of a body, at a given instant ($t_0$), you can then calculate (thus predict with certainty) any other position and velocity that the body will occupy, in any subsequent time, by solving the so-called \emph{equations of motion}. You can for instance determine its position $x_1$ and velocity $v_1$, at a subsequent time $t_1$, or its position $x_2$ and velocity $v_2$ at a further time $t_2$, and so on. This means that the body goes along a \emph{trajectory in space}, as time passes by, which is perfectly defined, i.e., a priori knowable with certainty. 

\begin{figure}[htbp] 
\begin{center} 
\includegraphics[width=10cm]{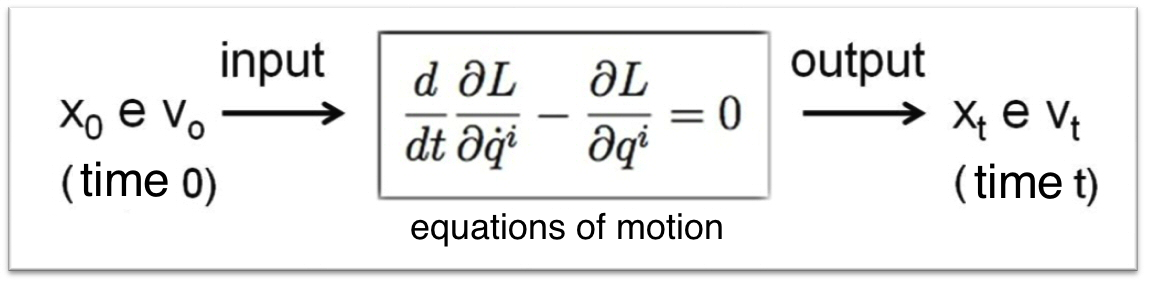} 
\caption{The equations of motion allow to predict every position and velocity of a macroscopic body, based on a precise input (the so-called \emph{initial condition}).} 
\label{Figure29}
\end{center} 
\vspace{-0.5cm}
\end{figure}

In other words, to solve the equations of motion is equivalent to predict with certainty any future position and velocity of the macroscopic body in question. I will not enter here into the details of these equations of motion, which depending on the physical systems can become quite complex. What is important to understand is that these equations are like a ``mechanical device,'' and when you feed such device with a precise input, formed by a position and a velocity, evaluated at a same instant of time, say at time $t=0$, the device will invariably provides you with outputs, corresponding to the positions and velocities at any other instant of time $t$, both in the future and in the past (see Figure~\ref{Figure29}). 

\begin{figure}[htbp] 
\begin{center} 
\includegraphics[width=3cm]{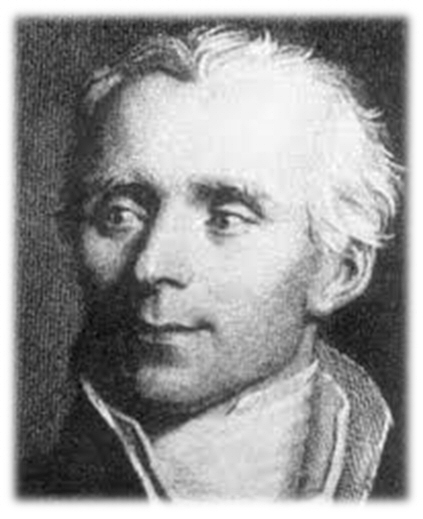} 
\caption{The French scholar Pierre-Simon de Laplace.} 
\label{Figure30}
\end{center} 
\vspace{-0.5cm}
\end{figure}

It is by considering this remarkable property of the equations of motion that the Frenchman \emph{Pierre-Simon de Laplace} (see Figure~\ref{Figure30}), towards the end of the eighteenth century, enunciated his famous principle of determinism, more or less in these words \cite{Laplace1825}:
\begin{displayquote}
\emph{If, at a certain moment, we would simultaneously know the position and the velocity of all bodies of the universe, then, in principle, we could predict their behavior at any other time, both in the past and in the future.} 
\end{displayquote}

For Laplace, the simultaneous knowledge of the position and velocity of all bodies in the universe was entirely possible, at least in principle. However, due to HUP, you know today that he was wrong, that a knowledge of this kind is absolutely unthinkable, and this not for a lack of information, or of an adequate technology. Indeed, if you remember, HUP does not allow one to jointly determine, with arbitrary precision, the position and velocity of a micro-entity, like an electron, let alone all those in the universe!

So, there is no way to insert in the equations of motion the input required and, consequently, the equations of motion are no longer able to provide you with the desired outputs. Accordingly, you are no longer able to predict, in no way, the position and velocity of the micro-entity under consideration, at any other instant of time $t$. And this means that the \emph{principle of determinism}, as it was enunciated by Laplace, is not valid for a microscopic entity. 

Failing the ability to determine, i.e., to predict with certainty, the future positions and velocities of a microscopic entity, what can you conclude on the basis of EPR-PA reality criterion? Well, it is very simple. The criterion tells you that the possibility to predict with certainty the position and/or velocity of a particle is equivalent to the reality, that is, to the existence, of the position and/or velocity of that corpuscle. But if the possibility of predicting these quantities fails, in the sense that they cannot be predicted, not even in principle, this means that they cannot be considered to be real, existing quantities. So, you are forced to conclude the following:
\begin{displayquote}
\emph{Microscopic particles (electrons, protons, neutrons, etc.) do not exist! In the sense that `they do not exist as particles', i.e., as entities that would be stably localized in space, thus possessing at any moment a well-defined position, velocity and energy.} 
\end{displayquote}

If I also say ``energy'' it is because, as is known, the energy of a material body is in general a function of both its velocity and position. And if these quantities are not actually existing, then the same must be true also for the energy. In short, the so-called ``microscopic particles," which particles are not, are \emph{non-spatial} entities! 

In other words, if a macroscopic body is able to possess, in every moment, a well-defined position, velocity and energy, a microscopic pseudo-corpuscle, instead, cannot possess in general such attributes. To say it with the thought provoking words of Diederik Aerts, we must surrender to the evidence that \cite{Aerts1999}:
\begin{displayquote}
\emph{Reality is not contained within space. Space is a momentaneous crystallization of a theatre for reality where the motions and interactions of the macroscopic material and energetic entities take place. But other entities -- like quantum entities for example -- ``take place" outside space, or -- and this would be another way of saying the same thing -- within a space that is not the three dimensional Euclidean space.} 
\end{displayquote}

This means that the three-dimensional space in which we live, with our physical body, macroscopic in nature, is only a small theater, which cannot contain all of our physical reality. Dimensionally speaking, reality is much bigger than that, and cannot be represented on such a small three-dimensional stage. So, there must be other ``stages'' out there, able to accommodate entities having a genuine non-spatial nature; entities whose spatiality is of a very different, non-ordinary kind. 

But if the microscopic entities generally do not have a position, what does this exactly mean? How can one understand the process through which a physicist, under certain experimental conditions, can observe the spatial position of an elementary entity? The answer given by Diederik Aerts is simple \cite{Aerts1999}: 
\begin{displayquote}
\emph{Quantum entities are not permanently present in space [...] when a quantum entity is detected in such a non-spatial state, it is `dragged' or `sucked up' into space by the detection system.} 
\end{displayquote}

Therefore, the spatial position (that is, the location) of a microscopic entity does not exist before the observational process, but is created by the very process of observation. But that's not all. To say it all, the spatial position of a microscopic entity does not even exist after the observational process. Indeed, it is a property of an \emph{ephemeral} nature \cite{Sassoli2011}.

At this point, perhaps you will ask: How can I understand the ephemerality and the incompatibility of quantum properties? Can I find deep analogies that can help me to better understand? Absolutely yes, and for this it is sufficient to love Italian spaghetti!

\section{The strange physics of spaghetti}

\begin{figure}[htbp] 
\begin{center} 
\includegraphics[width=3.5cm]{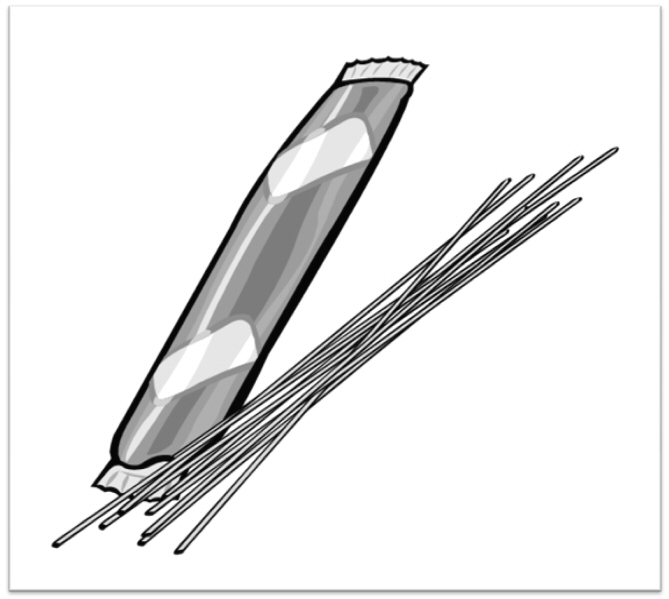} 
\caption{A typical package of Italian spaghetti.} 
\label{Figure31}
\end{center} 
\vspace{-0.5cm}
\end{figure}

With the final sentence of the previous section I wanted to say exactly what I said: that some of the quantum mysteries that you have explored so far can be clarified by studying the strange physics of spaghetti, and more precisely of raw spaghetti (see Figure~\ref{Figure31}).

For this, you will have to deal with the so-called (so to speak) \emph{left-handedness} of spaghetti. Let me explain what it is. So, the physical system (or physical entity) you want to study is an uncooked spaghetti (preferably of a good brand). The measuring instrument you are going to use, in order to perform your observational experiments, that is, your measurements, is formed by your two hands.

\begin{figure}[htbp] 
\begin{center} 
\includegraphics[width=6cm]{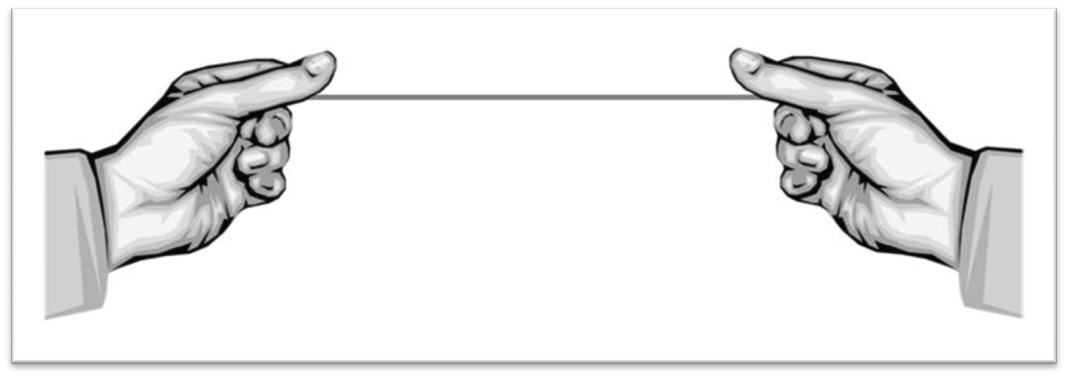} 
\caption{The procedure for observing the left-handedness of a spaghetti requires to initially grab the spaghetti with the two hands.} 
\label{Figure32}
\end{center} 
\vspace{-0.5cm}
\end{figure}

The property that you want to observe, as I said, is the left-handedness. I know, almost surely you have never heard of the left-handedness of a spaghetti, but I will now explain what it is, by telling you how it is measured/observed. You first have to grab the spaghetti with your two hands, as shown in Figure~\ref{Figure32}.

\begin{figure}[htbp] 
\begin{center} 
\includegraphics[width=6cm]{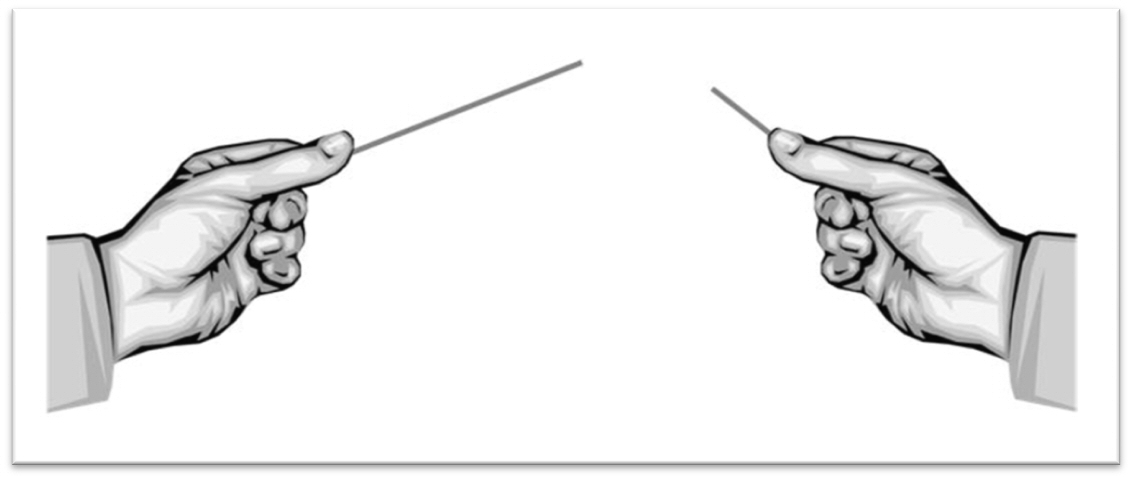} 
\caption{The test has confirmed the left-handedness property of the spaghetti. In other words, the observation was succesfull.} 
\label{Figure33}
\end{center} 
\vspace{-0.5cm}
\end{figure}

Then, you have to bend the spaghetti until it breaks; if the longest fragment remains in your left hand, then the left-handedness property is confirmed; otherwise, it is not confirmed. And if the spaghetti is already broken, you simply have to execute the test using the longest fragment. In Figure~\ref{Figure33}, you can see a possible result of such process. 

\begin{figure}[htbp] 
\begin{center} 
\includegraphics[width=6cm]{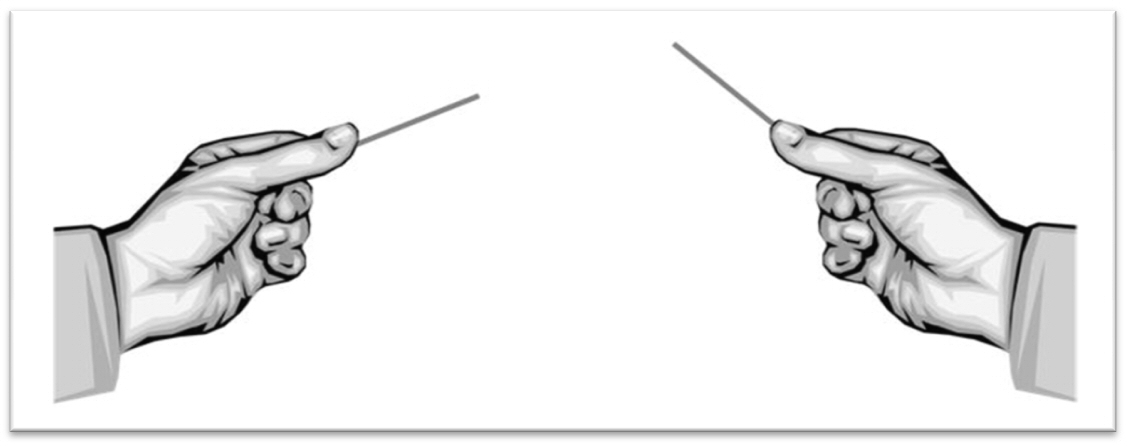} 
\caption{The test has not confirmed the left-handedness property of the spaghetti. In other words, the observation was not successful.} 
\label{Figure34}
\end{center} 
\vspace{-0.5cm}
\end{figure}

As you can observe, the test was successful, thus the left-handedness of the spaghetti was confirmed. In Figure~\ref{Figure34}, you can see the another possible result of the left-handedness observational process.

This time the test was not successful, and the left-handedness of the spaghetti was not confirmed. Instead, it is the inverse property of left-handedness, which is the property of \emph{right-handedness}, which was confirmed. 

Fine, but let me now consider another property of the spaghetti, which I will simply call the \emph{solidity}. So, the physical entity to be measured is once more an uncooked spaghetti. The measuring instrument is this time only one of your hands, combined with the floor of your kitchen. Again, to let you know what the solidity property is, all I have to do is to describe you the experimental observational protocol, which is as follows.

\begin{figure}[htbp] 
\begin{center} 
\includegraphics[width=6cm]{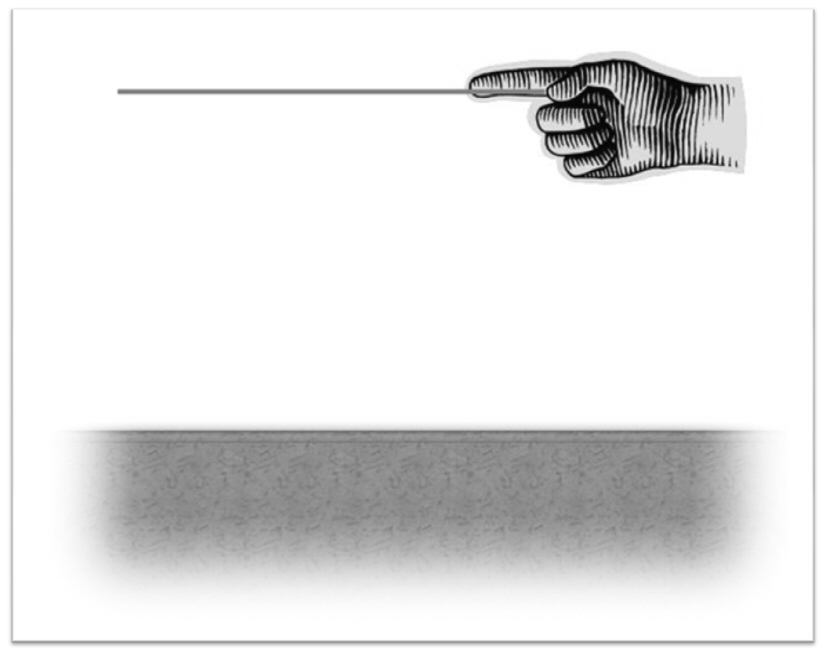} 
\caption{The procedure for observing the \emph{solidity} of a spaghetti requires to initially hold the spaghetti in one of your hands.} 
\label{Figure35}
\end{center} 
\vspace{-0.5cm}
\end{figure}

You first have to hold the spaghetti in your hand, as indicated in Figure~\ref{Figure35}. 

\begin{figure}[htbp] 
\begin{center} 
\includegraphics[width=6cm]{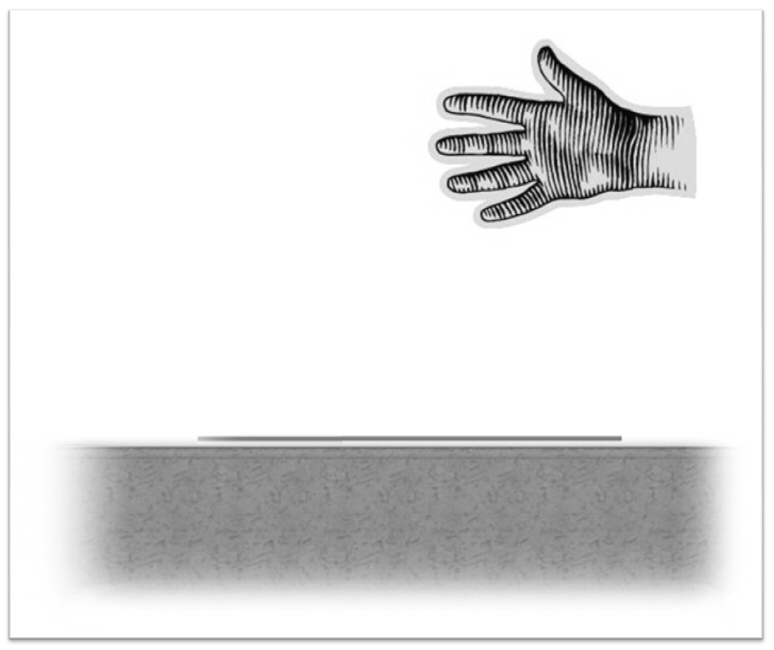} 
\caption{The successful outcome of the \emph{solidity} test.} 
\label{Figure36}
\end{center} 
\vspace{-0.5cm}
\end{figure}

\begin{figure}[htbp] 
\begin{center} 
\includegraphics[width=6cm]{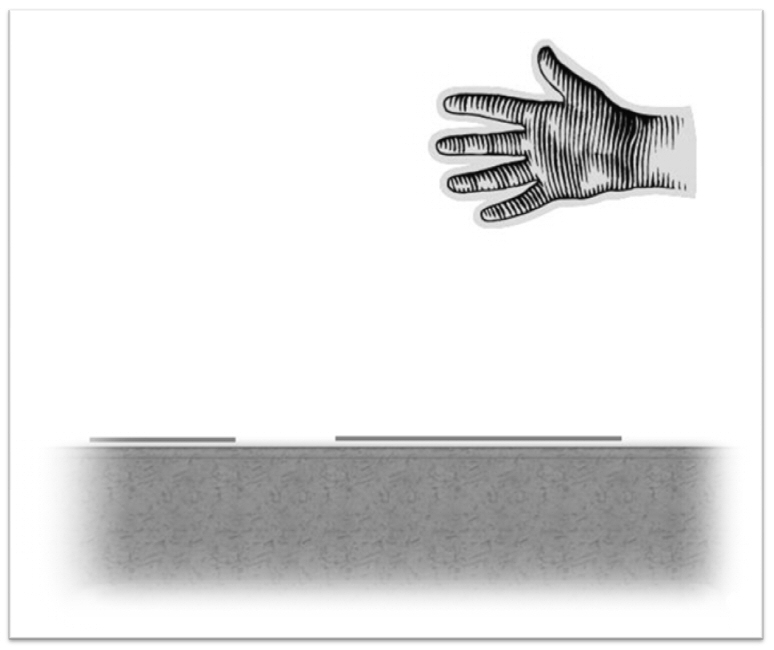} 
\caption{The unsuccessful outcome of the \emph{solidity} test, confirming the inverse property of \emph{fragility}. } 
\label{Figure37}
\end{center} 
\vspace{-0.5cm}
\end{figure}

Then, you have to let it fall from your hand to the floor, from a height of about one meter; if it doesn't break, as in Figure~\ref{Figure36}, the solidity property is confirmed; it it breaks, as in Figure~\ref{Figure37}, the solidity property is invalidated, whereas it is the inverse property of solidity, \emph{fragility}, which is confirmed. And if the spaghetti is already broken, you simply do the experiment using the longest fragment. 

Very well, if you have thought carefully, you may have understood that: 
\begin{displayquote}
\emph{Left-handedness and solidity are ephemeral properties}.
\end{displayquote}

Indeed, consider the case of the left-handedness. Suppose that you have just made the observation and that the test was successful. At that precise moment, when you are still holding the two fragments of spaghetti in your hands, you can certainly say that the spaghetti \emph{actually} possesses the left-handedness property. But as soon as you let go of the two fragments, that same left-handedness goes back to being a property which is only \emph{potential}. 

In fact, once the \emph{relation} between the fragments of the spaghetti and the two hands of the experimenter is lost, it is no longer possible to affirm that the spaghetti is left-handed \cite{Sassoli2015}. This because to observe again the left-handedness, you have to repeat the test, using the longer fragment, but nothing a priori guarantees you that its outcome will be again successful. 

I hope you will appreciate the fundamental difference between the observation of a property such as the burnability of a wooden cube, and the left-handedness of a spaghetti. For the burnability, you were perfectly able to predict the outcome of the test, without any need to perform it. On the other hand, for left-handedness you are no longer in such situation.

\begin{figure}[htbp] 
\begin{center} 
\includegraphics[width=5cm]{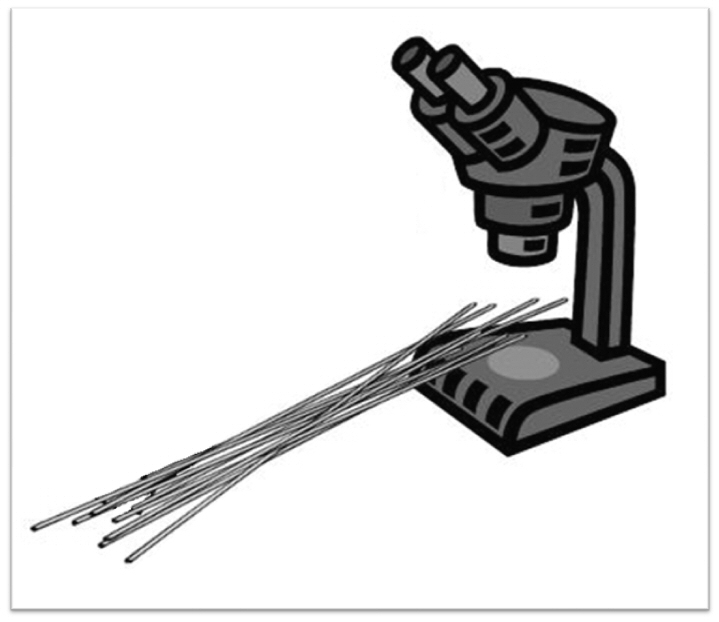} 
\caption{Studying spaghetti under a microscope does not make it easier to predict the outcome of the left-handedness (or solidity) test. } 
\label{Figure38}
\end{center} 
\vspace{-0.5cm}
\end{figure}

Of course, you could argue that to make a reliable prediction you need to have all the necessary information, and for this carefully study in advance, maybe under a microscope (see Figure~\ref{Figure38}), all the characteristics of the spaghetti in question, perhaps also asking precise information from the manufacturer about the manufacturing method. 

But will this really help you to predict in advance the outcome of the left-handedness (or solidity) test? If you attentively consider the way in which the left-handedness property (or the inverse property of right-handedness) is tested, you will easily convince yourself that even iwith a complete knowledge of the spaghetti, up to the level of its molecular structure, you will never be able to predict with certainty the outcome of the observational test. 

This is not because you would lack some essential information about the spaghetti as such (that is, about its actual \emph{state}), but because you remain totally clueless about how exactly the process of the observation of the property will took place. 

The outcome of the test will in fact depend on a number of variables that remain hidden to us, in the sense that are totally \emph{outside of our control}, such as the subtle vibrations of your hands while you act on the spaghetti, its specific orientation, the variable pressure exerted by your fingers, the rapidity with which you bend it in order to break it, and so forth. 

It is the combination of all these \emph{hidden variables}, and those associated with the state of the spaghetti, which will determine, in an extremely complex way, in which point(s) the spaghetti is going to break, thus producing the final outcome of the test. 

In other words, despite possibly having a complete knowledge of the state of the spaghetti, you will have no way to predict the effects of the innumerable fluctuations in the interaction between the spaghetti and the instrument of observation, made by your two hands; fluctuations that ultimately will determine the exact breaking point(s) of the spaghetti, and therefore either its left-handedness or right-handedness. 

The logical conclusion of all this is that the property of the spaghetti of being left-handed or right-handed cannot be predicted in advance with certainty, also when you possess a complete knowledge of the state of the spaghetti. And according to the EPR-PA criterion of reality, this means that these properties cannot be considered to be genuine elements of reality, i.e., they do not exist. Or, rather, their existence is only \emph{potential}, in the sense that although they do not exist (they are not actual) in a given moment, they may nevertheless exist (be actualized) at a later instant. This is exactly what can happen during an observational process: 
\begin{displayquote}
\emph{The left-handedness (or solidity) property is not `discovered' during its observational test, but possibly literally `created' by it. Before the test, it was not existing, but by means of the test it can be brought into existence, although only in an ephemeral way.} 
\end{displayquote}

To sum up: left-handedness and solidity are both ephemeral properties, in the sense that they are potential properties that are possibly created/actualized during an observation, by the observation itself, in a way that cannot be predicted in advance. Also, they cease to be actual at the precise moment when the specific relation between the measuring instrument and the physical entity is severed.

Left-handedness and solidity, however, are also properties that are mutually incompatible. In fact, the observation of the left-handedness considerably increases the probability that a subsequent test of the solidity property will give a successful outcome, as is clear from the fact that the shortest the spaghetti the less easily it will break, when falling to the floor. 

This means that, in general, performing first the test of left-handedness, then afterwards the test of solidity, for example on a large number of different spaghetti, the statistics of outcomes you obtain will differ significantly from that obtained by first performing the solidity test and then the left-handedness one. This paradigmatic example of the spaghetti \cite{Sassoli2011} reveals in an incontrovertible way that: 
\begin{displayquote}
\emph{Incompatibility and ephemerality are independent notions}.
\end{displayquote}

Indeed, as Aerts' piece of wood example shows \cite{Aerts1982}, two properties can be incompatible and nevertheless stably exist, at the same time; but as the spaghetti example also shows, two properties can be incompatible and only ephemerally exist, when actualized by their experimental test. 

This indicates that the ephemeral character of a property has more to do with the way the property itself is defined (i.e., the way it is tested, in a practical way) than the fact that it may or not entertain incompatibility relations with other properties. 

The above remark is particularly relevant in view of the fact that in our reasoning to deduce the non-spatiality of microscopic entities, the HUP (the existence of a relation of incompatibility between position and velocity) was used as a main ingredient. 

Therefore, one could be tempted to conclude that it is the very existence of such an experimental incompatibility which is at the origin of the observed non-spatiality of the microscopic entities. 

Considering the piece of wood example, we see however that incompatibility is not a sufficient condition for non-spatiality, and considering the spaghetti example, we also see that incompatibility is neither a necessary condition for it, as is clear from the fact that the ephemeral character of the left-handedness and solidity properties is built-in in the very definition of them, independently of the compatible or incompatible nature of their relation with other properties. 

That being said, I hope I have not lost you in all these conceptual subtleties. What is really important to highlight here is that the left handedness and solidity of a spaghetti, like the position and velocity of a microscopic entity, are \emph{non-ordinary} properties, in the sense that they are \emph{non-classical} properties, non-spatiality being just an aspect of such non-classicality. 

But perhaps you are now wondering: what exactly are classical properties? Well, simply, classical properties are properties obeying the so-called \emph{classical prejudice} \cite{Piron1998}, stating that: 
\begin{displayquote}
Classical prejudice: \emph{the outcome of an observational test is always a priori certain (predetermined), i.e., always predictable in advance, at least in principle}. 
\end{displayquote}

But the classical prejudice has a very limited validity, being based on the wrong assumption that the interaction between the instrument of observation and the observed system/entity always takes place in a predeterminable way. 

The observational tests of burnability and floatability of the wooden cube are certainly in accordance with the classical prejudice. But the observational tests of left-handedness and solidity of the raw spaghetti certainly invalidate the classical prejudice, in the same way it is invalidated by the observation of the position and velocity of a microscopic entity.

Very well. Let me now summarize the most important points of our investigation:
\begin{itemize}
\item We have seen that HUP expresses the experimental incompatibility of certain properties associated with microscopic entities, like position and velocity.
\item We have also highlighted that experimental incompatibility is a widespread phenomenon, which manifests also with macroscopic bodies, and not only with microscopic ones.
\item Moreover, and contrary to what one might initially believe, we have shown that it is perfectly possible to jointly test incompatible properties, by means of the so-called product tests.
\item We have then highlighted the content of the EPR reality criterion, and its more complete EPR-PA version by Constantin Piron and Diederik Aerts, affirming that existence and predictability are intimately related notions (in the sense that, in ultimate analysis, a property is a state of prediction).
\item Next, using in combination the HUP and the EPR-PA criterion, we have deduced the non-spatiality of microscopic entities, showing that to have a position is an ephemeral property of a microscopic entity, not stably possessed by it.
\item We have then seen that ephemerality can also manifest in macroscopic bodies, and that incompatibility and ephemerality are independent notions.
\item Finally, we realized that wooden cubes and uncooked spaghetti can be of great help in understanding (and in part demystifying) some of the mysteries of quantum physics. 
\end{itemize}

Of course, much more should be said to elucidate all these conceptually profound and subtle aspects of our physical theories. In particular, much should be added concerning the puzzling non-spatiality of quantum entities, often indicated by physicists with the less appropriate term of \emph{non-locality} (as the latter implicitly suggests that the entity would still remain stably present within our spatial theater, although in a sort of spatially widespread condition).

\section{The friendship space}

I would like to conclude this presentation with a metaphor proposed in 1990 by Diederik Aerts \cite{Aerts1990}, as an attempt to invent a world of entities where their spatial condition emerges from a different underlying reality, whose spatiality is of a different kind. This world of entities is, as I'm going to explain, a world living within a space of friendship. 

More precisely, the entities considered are we human beings in a distant future, and the world of entities is our terrestrial human society. The interaction taken into account is the one of \emph{friendship}, and the hypothesis is that the human society was able to survive by managing to totally eliminate \emph{enmity} (negative friendship) and by making friendship always something \emph{reciprocal}.
Let me explain what this means in more precise terms. If you denote by ${\rm ad}(X,Y)$ the function that determines the \emph{affective distance} that person $X$ feels for person $Y$, and if ${\rm ad}(Y,X)$ denotes the affective distance that person $Y$ feels for person $X$, then reciprocity simply means that these two distances are identical or, to say it in more technical terms, that the ``ad'' function is \emph{symmetrical}: 
\begin{equation}
{\rm ad}(X,Y) = {\rm ad}(Y,X).
\end{equation}

The absence of enmity, on the other hand, means that the function ad is always positive, as it should be the case for a distance worth of the name:
\begin{equation}
{\rm ad}(X,Y) \geq 0.
\end{equation}

let me consider now the \emph{ordinary physical space} in which we humans live today.
In this space there are people, for example, a boy, who I will call $X$, and a girl, who will call $Y$.

Between these two guys there is a \emph{physical distance}, which I will denote ``pd,'' which is the distance usually considered between the ordinary physical spatial objects, also a symmetrical function: 
\begin{equation}
{\rm pd}(X,Y) = {\rm pd}(Y,X).
\end{equation}

The boy $X$ and the girl $Y$ do not only exist in the ordinary physical space, but also in the space of friendship, and in this other space $X$ and $Y$ are not separated by a physical distance ``pd,'' but by an affective distance ``af,'' which for example we can consider to be much smaller than the physical one, as is clear that the mutual friendship between two people does not depend on how distant they are in physical terms.

\begin{figure}[htbp] 
\begin{center} 
\includegraphics[width=9cm]{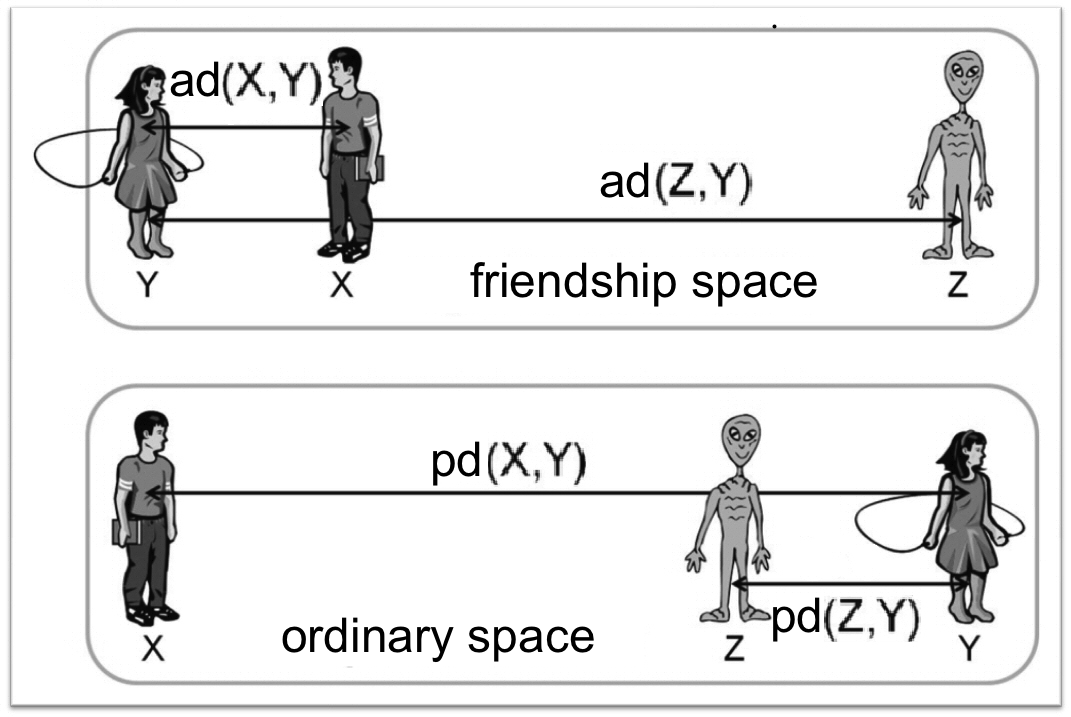} 
\caption{Distances in the physical space and in the friendship space follow different logics: nearby objects in the physical space can be pretty much far away in the friendship space, and vice versa.} 
\label{Figure39}
\end{center} 
\vspace{-0.5cm}
\end{figure}

Consider now a third individual, who I will call $Z$, whose physical distance with $Y$ is smaller than that between $X$ and $Y$, and whose distance in the friendship space is larger than that between $X$ and $Y$, for reasons that you can easily guess by looking at Figure~\ref{Figure39}. 

Very well, up to here I have simply pointed out that the distances in the physical space and in the friendship space are not necessarily in correspondence with each other. Consider now the fact that in our human society, as time passes by, different subgroups of people will \emph{emerge}, bound by specific affinities, which for simplicity I will consider to only be \emph{affinities of an affective kind}. This emergence is responsible for a \emph{structuring of the friendship interaction}. 

\begin{figure}[htbp] 
\begin{center} 
\includegraphics[width=9cm]{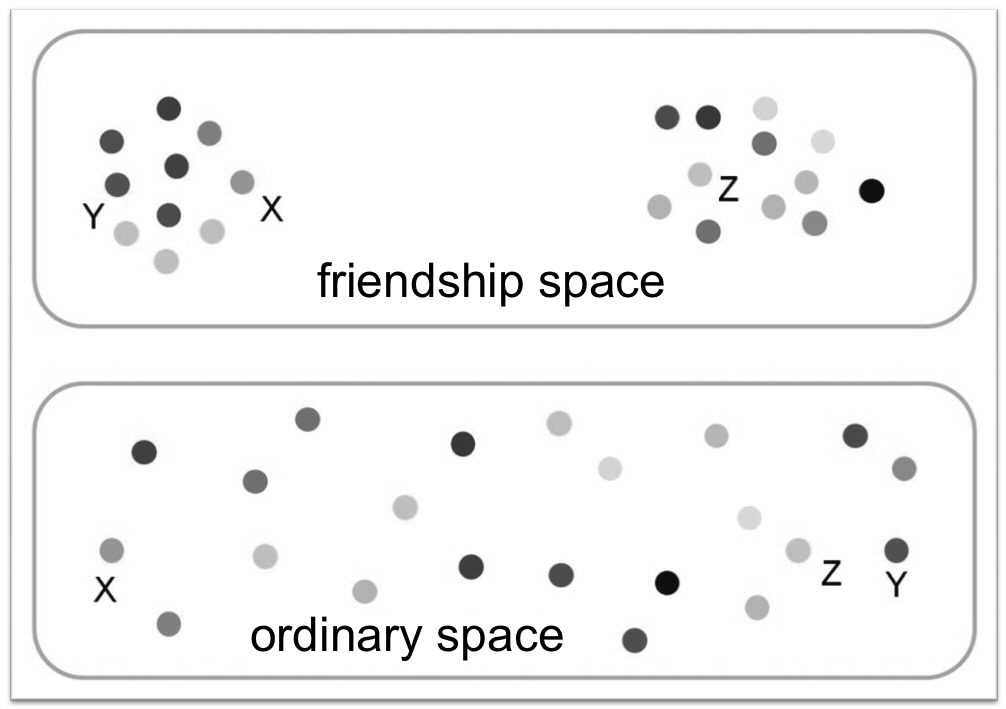} 
\caption{The same individuals are represented in the friendship space (above) and in the ordinary physical space (below).} 
\label{Figure40}
\end{center} 
\vspace{-0.5cm}
\end{figure}

Try to observe this phenomenon from the double perspective of the physical space and friendship space. In Figure~\ref{Figure40}, different persons (including $X$, $Y$ and $Z$) are represented (for simplicity as simple dots). As you can see, although the different individuals are rather scattered in the ordinary physical space, they present themselves in a much more organized way in the friendship space.

\begin{figure}[htbp] 
\begin{center} 
\includegraphics[width=9cm]{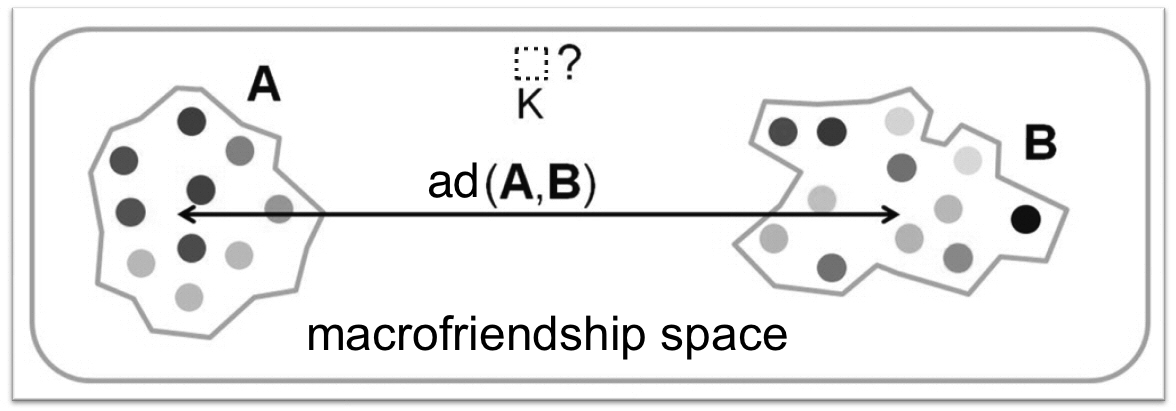} 
\caption{Two macrostructures, $A$ and $B$, in the \emph{space of macrofriendship}, separated by a distance ${\rm ad}(A,B)$. The individual entity $K$, not belonging to $A$ or $B$, is in a state of \emph{superposition} with respect to these macrostructures, hence does not belong to the macrofriendship space.} 
\label{Figure41}
\end{center} 
\vspace{-0.5cm}
\end{figure}

More precisely, as made explicit in Figure~\ref{Figure41}, people end up organizing themselves into \emph{macrostructures}. For simplicity, in Figure~\ref{Figure41} I have only evidenced two of them, denoting them $A$ and $B$, which are located at a certain (affective) distance ${\rm ad}(A,B)$ from each other.

So, in the ``old'' physical space of the surface of planet earth, people live pretty much mixed together, but in the ``new'' structured space of friendship they are organized within specific \emph{macrostructures}. To fix ideas, we can think of families, associations, interest groups, sects, etc. In other words, as time goes on, the friendship space will, little by little, becomes a perfectly structured \emph{macrofriendship space}.
Consider now an additional individual $K$, and suppose that this individual, although at a given moent s/he is present in the ordinary physical space, s/he has not yet established a specific relationship with one of the macrostructures present in the macrofriendship space. 

Suppose that after a very long time, humans of the future have totally forgotten about their original Euclidean space, associated to the surface of planet earth, as well as about the first version of their friendship space, when it was not yet structured into well-defined (affective) macro structures. 

Then, in the fully structured space of macrofriendship, a single affectively isolated individual cannot have any localization, that is, a specific position, as having a well-defined localization in the macrofriendship space means to belong to a specific structure of affinity, in the present case either $A$ or $B$.

The individual $K$, from the viewpoint of the macrofriendship space, is therefore a typical \emph{non-spatial} entity, not present in actual terms in that space (see Figure~\ref{Figure41}). 

\begin{figure}[htbp] 
\begin{center} 
\includegraphics[width=9cm]{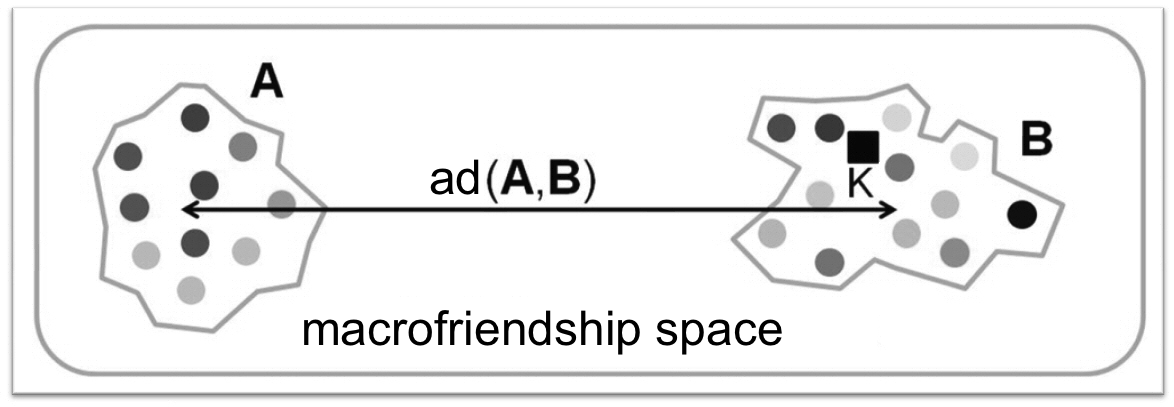} 
\caption{The individual $K$, by choosing which affective macrostructure to belong to, passes from a \emph{non-spatial superposition state} to a ``collapsed'' \emph{localized spatial state}, relative to the space of macrofriendship.} 
\label{Figure42}
\end{center} 
\vspace{-0.5cm}
\end{figure}

Suppose however that as a result of a (non-spatial) interaction with the existing macrostructures, it comes the time when, for reasons that we do not need to specify here, $K$ decides (or is forced) to choose to belong to either $A$ or $B$. Before that this happens, we can say that $K$ is in a quantum-like \emph{superposition state}, with respect to these two possibilities, and that at the exact moment s/he chooses to which macrostructure s/he belongs, $K$ suddenly acquires (collapses to) a specific location in the macro-friendship space, becoming for example an integral part of the macrostructure $B$ (see Figure~\ref{Figure42}).


It is interesting to note that this process, during which a specific localization for the elementary entity $K$ is suddenly created, in the macrofriendship space, through its interaction with the macrostructures $A$ and $B$, is a typical \emph{creation process}, which is reminiscent of the creation of a position for an electron, when it interacts with measuring instruments, which are precisely structures of a macroscopic kind, formed by a huge number of elementary microscopic entities, organized together. 

For sure, the above is just a metaphor \cite{Aerts1990}, though a very profound and enlightening one, which is certainly good to let it settle in your mind and on which it can be advantageous to further meditate. In fact, I believe the time has come to conclude this already quite long exposition.

\section{Reading suggestions}

If you have been interested in what you have read, and would like to deepen your reflection, here are some further reading suggestions.

Let me start with some texts that are readable also by those who are not experts in quantum physics. I will also then indicate a couple of articles for readers with some technical knowledge, wanting to dig into the formalism behind some of my explanations.

The fact that quantum (or quantum-like) measurements as also processes of \emph{creation}, and not only as processes of \emph{discovery}, is sometimes referred to as an \emph{observer effect}, where the term generally refers to the possibility that an observation may affect the properties of what is observed. Examples and illustrations of such observer (creation) effect can be found in \cite{Sassoli2013,Sassoli2013b,Sassoli2015,Sassoli2018}.

The left-handedness (or solidity) property of spaghetti subtends a possible interpretation of quantum measurements called the \emph{hidden-measurement interpretation} (HMI), which recently gave rise to a promising completion of the quantum formalism, known as the \emph{extended Bloch representation} (EBR) of quantum mechanics. A highly accessible introduction to the HMI and the EBR can be found in \cite{AertsSassolideBianchi2015b,AertsSassolideBianchi2017,Sassoli2017}; see also the video \cite{SassoliSassoli2015}, where some nice computer animations of the unfolding of quantum measurements with two, three and four possible outcomes can be found. 

For those readers who have fully mastered the quantum formalism, a more technical reading about the HMI and the EBR, containing all the mathematical details, can be found in \cite{AertsSassolideBianchi2014,AertsSassoli2016b}. 

Finally, regarding the possibility of considering the micro-entities as non-spatial entities, it is important to say that there are many different ways to reach such conclusion. In addition to my reasoning using the equations of motion, one can for instance analyze the remarkable experiments conducted in \emph{neutron interferometry}, for instance those observing the $4\pi$-periodicity of a neutron's spinor wave function, which one can be made to interfere with itself \cite{Rauchetal1975,Werneretal1975,Aerts1999,SassolideBianchi2017}. And speaking of \emph{spin}, it is possible to show that \emph{spin eigenstates} cannot in general be associated with directions in the \emph{Euclidean space}, but only with generalized directions in the \emph{Blochean space} \cite{AertsSassolideBianchi2017c}. Non-spatiality can also be deduced by considering the \emph{permanence time} of micro-entities in certain regions of space, which again can be shown to be incompatible with the very notion of a spatial trajectory \cite{Sassoli2012}.

To conclude, let me also point out a fascinating interpretation of quantum mechanics that was introduced by Diederik Aerts some years ago, and is currently under development, known as the \emph{conceptuality interpretation}. According to it, quantum entities would be non-spatial simply because they would be \emph{conceptual (abstract) entities}, interacting among them and with the measuring apparatuses in ways that are analogous to how human concepts combine with each other in our linguistic constructions and interact with human minds. This not because human concepts and the microscopic physical entities would be the same kind of entities, but because they would share the same \emph{conceptual nature}, similarly to how \emph{sound waves} and \emph{electromagnetic waves}, although very different entities, can nevertheless share the same \emph{ondulatory nature}. 

A good place to start, to learn more about this truly fascinating interpretation, is the recent review article \cite{Aertsetal2017}, and of course the references cited therein. A more concise video version of the article is also available on YouTube \cite{Sassoli2017v}, with the presentation I gave at the Symposium ``Worlds of Entanglement,'' organized by the Centre Leo Apostel for Interdisciplinary Studies and which took place at the Free University of Brussels (VUB), on September 29–30, 2017.

\end{document}